\documentclass[12pt,eqno]{article}
\usepackage{amsmath,amssymb,epsfig}
\numberwithin{equation}{section}

\def\ignore#1{{}}
 
\newcommand{\alp}{\alpha}
\newcommand{\bt}{\beta}

\newcommand{\dlt}{\delta}
\newcommand{\Dlt}{\Delta}
\newcommand{\ep}{\epsilon}
\newcommand{\vep}{\varepsilon}
\newcommand{\tht}{\theta}

\newcommand{\vtht}{\vartheta}

\newcommand{\lmd}{\lambda}
\newcommand{\Lmd}{\Lambda}
\newcommand{\sgm}{\sigma}
\newcommand{\Sgm}{\Sigma}

\newcommand{\Omg}{\Omega}

\newcommand{\be}{\begin{equation}}
\newcommand{\ee}{\end{equation}}
\newcommand{\bea}{\begin{eqnarray}}
\newcommand{\eea}{\end{eqnarray}}
\newcommand{\eql}{\!\!\!&=\!\!\!&}

\newcommand{\defa}{\!\!\!&\equiv\!\!\!&}

\newcommand{\simlt}{\stackrel{<}{{}_\sim}}
\newcommand{\sima}{\!\!\!&\simeq\!\!\!&}

\newcommand{\tl}[1]{\tilde{#1}}
\newcommand{\bdm}[1]{{\mbox{\boldmath $#1$}}}
\newcommand{\tr}{{\rm tr}\,}

\newcommand{\diag}{{\rm diag}\,}
\newcommand{\der}{\partial}
\newcommand{\dr}{\!\!d}
\newcommand{\hc}{{\rm h.c.}}
\newcommand{\ie}{{\it i.e.}}

\newcommand{\vev}[1]{\langle #1 \rangle}

\newcommand{\brkt}[1]{\left( #1 \right)}
\newcommand{\brc}[1]{\left\{ #1 \right\}}
\newcommand{\sbk}[1]{\left[ #1 \right]}
\newcommand{\abs}[1]{\left| #1 \right|}

\renewcommand{\Re}{{\rm Re}\,}

\newcommand{\cD}{{\cal D}}
\newcommand{\cF}{{\cal F}}

\newcommand{\cH}{{\cal H}}

\newcommand{\cL}{{\cal L}}

\newcommand{\cN}{{\cal N}}
\newcommand{\cO}{{\cal O}}

\newcommand{\cQ}{{\cal Q}}

\newcommand{\cS}{{\cal S}}

\newcommand{\cV}{{\cal V}}
\newcommand{\cW}{{\cal W}}

\renewcommand{\ge}[2]{e_{#1}^{\;\;#2}}

\newcommand{\gey}{\ge{y}{4}}
\newcommand{\SUu}{SU(2)_{\mbox{\scriptsize $\bdm{U}$}}}

\newcommand{\NP}[1]{{\it Nucl.~Phys.}~{\bf #1}}
\newcommand{\PL}[1]{{\it Phys.~Lett.}~{\bf #1}}

\newcommand{\PR}[1]{{\it Phys.~Rev.}~{\bf #1}}
\newcommand{\PRL}[1]{{\it Phys.~Rev.~Lett.}~{\bf #1}}
\newcommand{\PTP}[1]{{\it Prog.~Theor.~Phys.}~{\bf #1}}

\newcommand{\JH}[1]{{\it JHEP}~{\bf #1}}

\begin{document}

\begin{titlepage}
\null
\begin{flushright}
YITP-08-64
\\
RIKEN-TH-134
\\
July, 2008
\end{flushright}

\vskip 2cm
\begin{center}
\baselineskip 0.8cm
{\LARGE \bf 
Flavor structure with multi moduli in 5D supergravity
}

\lineskip .75em
\vskip 2.5cm

\normalsize

{\large\bf Hiroyuki Abe}${}^1\!${\def\thefootnote{\fnsymbol{footnote}}
\footnote[1]{\it e-mail address: abe@yukawa.kyoto-u.ac.jp}}
{\large\bf and Yutaka Sakamura}${}^2\!${\def\thefootnote{\fnsymbol{footnote}}
\footnote[2]{\it e-mail address: sakamura@riken.jp}}

\vskip 1.5em

${}^1${\it Yukawa institute for theoretical physics, Kyoto University, \\ 
Kyoto 606-8502, Japan}

\vskip 1.0em

${}^2${\it RIKEN, Wako, Saitama 351-0198, Japan}

\vspace{18mm}

{\bf Abstract}\\[5mm]
{\parbox{13cm}{\hspace{5mm} \small
We study 5-dimensional supergravity on $S^1/Z_2$ with a 
physical $Z_2$-odd vector multiplet, which yields an additional 
modulus other than the radion. 
We derive 4-dimensional effective theory and find additional terms 
in the K\"{a}hler potential that are peculiar to the multi moduli case. 
Such terms can avoid tachyonic soft scalar masses at tree-level, 
which are problematic in the single modulus case. 
We also show that the flavor structure of the soft terms are different 
from that in the single modulus case when hierarchical Yukawa couplings 
are generated by wavefunction localization in the fifth dimension. 
We present a concrete model that stabilizes the moduli at a 
supersymmetry breaking Minkowski minimum, 
and show the low energy sparticle spectrum. 
}}

\end{center}

\end{titlepage}


\section{Introduction}
Supersymmetry (SUSY) is one of the most promising candidates 
for the physics beyond the standard model. 
It solves the gauge hierarchy problem 
in a sense that it stabilizes the large hierarchy 
between the Planck scale~$M_{\rm Pl}\sim 10^{19}$~GeV 
and the electroweak scale~$M_{\rm weak}\sim 10^2$~GeV 
under radiative corrections. 
Especially the minimal supersymmetric standard model (MSSM) 
predicts that the three gauge couplings in the standard model 
are unified around $M_{\rm GUT}\simeq 2\times 10^{16}$~GeV, which suggests 
the grand unified theory (GUT). 
It also has a candidate 
for the cold dark matter if the R-parity forbids 
decays of the lightest SUSY particle. 
Besides, the existence of SUSY is predicted by 
the superstring theory, which is a known consistent theory 
of the quantum gravity, together with extra spatial dimensions 
other than the observed four-dimensional (4D) spacetime. 

Since no SUSY particles have not been observed yet, 
SUSY must be spontaneously broken above $M_{\rm weak}$. 
Such effects in the visible sector 
are summarized by the soft SUSY breaking parameters. 
Arbitrary values are not allowed for these soft parameters 
because they are severely constrained from 
the experimental results for the flavor changing processes. 
This is the so-called SUSY flavor problem. 

Models with the extra dimensions 
have been investigated in a large number  
of articles since the possibility was pointed out 
that the gauge hierarchy problem is solved 
by the introduction of the extra dimensions~\cite{ADD,RS}.  
The extra dimensions can play many other important roles 
even in the case that the gauge hierarchy problem is solved 
by SUSY. 
For instance, they can generate the hierarchy among 
quarks and leptons by localized wave functions 
in the extra dimensions~\cite{Arkani-Schmaltz}. 
In fact, SUSY extra-dimensional models have attracted 
much attention as a candidate for the physics 
beyond the standard model. 

In many works on the extra-dimensional models, 
the size and the shape of the extra-dimensional compact space 
are treated as given parameters of the models. 
However they should be considered as 
dynamical variables called the {\it moduli} 
and should be stabilized to some finite values by the dynamics. 
In order to discuss the moduli stabilization 
in SUSY extra-dimensional models, 
we have to work in the context of supergravity (SUGRA). 
The moduli belong to chiral multiplets 
when a low-energy effective theory below 
the compactification scale is described as 4D SUGRA. 
Vacuum expectation values (VEVs) of the moduli determine 
quantities in the 4D effective theory such as $M_{\rm Pl}$, 
the gauge and Yukawa couplings. 
The moduli stabilization is also quite relevant to 
the soft SUSY breaking terms because moduli multiplets 
generically couple to the visible sector in the effective theory, 
and their $F$-terms are determined by the scalar potential 
that stabilizes the moduli themselves. 

Five-dimensional (5D) SUGRA compactified on 
an orbifold~$S^1/Z_2$ is the simplest setup 
for SUSY extra-dimensional models and has many interesting 
features which are common among them. 
Furthermore it has an off-shell description that makes 
the SUSY structure manifest and also 
allows us to deal with the actions in the bulk and 
at the orbifold boundaries 
independently~\cite{Zucker,Kugo-Ohashi,Kugo:2002js}. 
There is another advantage of 5D SUGRA models, 
that is, the explicit calculability of the 4D effective theory. 
This is in contrast to the superstring models whose 
4D effective theories are complicated and difficult 
to be derived explicitly. 
On the other hand, the 4D effective theory 
of 5D SUGRA models can be easily calculated by 
a method which we call the 
{\it off-shell dimensional reduction}~\cite{Abe:2006eg}.  
This is based on the $N=1$ superspace\footnote{
$N=1$ SUSY denotes four supercharges in this paper. }  
description~\cite{Paccetti Correia:2004ri,Abe:2004}
of the 5D conformal SUGRA and developed 
in subsequent studies~\cite{ASworks,Correia:2006pj}. 
This method has the advantage that the $N=1$ off-shell SUSY structure 
is kept during the derivation of the 4D effective theory. 
Furthermore this method can be applied to general 5D SUGRA models. 
For example, we analysed some class of 5D SUGRA models 
by using this method,  
including the SUSY extension of 
the Randall-Sundrum model~\cite{SUSY-RS} 
and the 5D heterotic M theory~\cite{5D_Mtheory} 
as special limits of the parameters~\cite{Abe:2007}. 
The effective theory approach makes it easy 
to discuss the moduli stabilization in these models. 

In 5D models, there is only one modulus that originates from 
the extra dimension, that is, the radion. 
Most works dealing with 5D SUSY models assume that 
it belongs to a chiral multiplet as the real part of 
its scalar component (see, e.g, Ref.~\cite{Luty-Sundrum}).  
However this is only true for models that have 
no zero-mode for the scalar component 
of a 5D vector multiplet. 
If there exist such zero-modes in the 4D effective 
theory, the radion must be mixed with them 
to form a chiral multiplet. 
In this sense, those zero-modes are on equal footing 
with the radion,\footnote{
In fact, the radion is regarded as a zero-mode for 
the scalar component of the graviphoton vector multiplet 
in the off-shell 5D SUGRA description.  
}
and thus we also call them the moduli in this paper. 
They actually correspond to the shape moduli of 
the internal compact manifold when the 5D SUGRA model 
is the effective theory of the heterotic M theory 
compactified on a Calabi-Yau 3-fold~\cite{5D_Mtheory}. 
In this paper, we will consider 5D SUGRA with {\it multi moduli}, 
and investigate its effective theory, 
focusing on the flavor structure of the soft SUSY breaking 
parameters. 
Such flavor structure was studied in Ref.~\cite{single_modulus} 
in the single modulus case. 
The main results there are the following. 
First, the soft scalar masses tend to be 
tachyonic at tree-level. 
This problem can be solved by sequestering the SUSY breaking sector 
from the visible sector because the quantum effects dominate 
over the tree-level contribution in such a case. 
However, generation of the Yukawa hierarchy 
and the sequestering of the SUSY breaking sector 
cannot be achieved simultaneously 
if they are both realized by localized wave functions 
in the fifth dimension. 

The essential difference from the single modulus case 
appears in the K\"{a}hler potential in the 4D effective action. 
This difference comes from contributions mediated by 
the $Z_2$-odd $N=1$ vector multiplets. 
Although they have no zero-modes and 
do not appear in the effective theory, 
the effective K\"{a}hler potential is modified 
after they are integrated out. 
We show that this contribution can save the problems 
in the single modulus case, which are mentioned above. 

In the multi moduli case, it is generically difficult to 
explicitly calculate wave functions in the fifth dimension 
for the 4D modes because the mode equations are complicated 
coupled equations. 
This makes it hard to derive the 4D effective theory 
by the conventional Kaluza-Klein (KK) dimensional reduction. 
In the off-shell dimensional reduction, however, the 4D effective theory 
is obtained without calculating the wave functions explicitly. 
This is also one of the advantages of our method. 

The moduli stabilization and SUSY breaking are discussed 
in a specific model.   
We consider a situation where SUSY is broken by the $F$-term of one 
chiral multiplet~$X$ in the effective theory. 
By utilizing a technique developed in Ref.~\cite{AHKO}, 
we find a vacuum where the moduli are stabilized properly 
and the $F$-term of $X$ is certainly a dominant source of SUSY breaking. 
We also examine the flavor structure of the soft SUSY breaking parameters 
at $M_{\rm weak}$ in this model 
by using the renormalization group equations (RGEs). 

The paper is organized as follows. 
In Sec.~\ref{derive_4DEFT}, we give a brief review of 
our method to derive 4D effective theory of 5D SUGRA on $S^1/Z_2$. 
In Sec.~\ref{soft_terms}, we discuss generic properties of 
the soft SUSY breaking parameters in the multi moduli case. 
In Sec.~\ref{moduli_stb}, the moduli stabilization and SUSY breaking 
are discussed in a specific model, and the soft SUSY breaking parameters 
at $M_{\rm weak}$ are evaluated 
by the numerical calculation. 
Sec.~\ref{summary} is devoted to the summary. 
In Appendix~\ref{Z2odd_mul}, we comment on how the $Z_2$-odd part 
of the 5D Weyl multiplet appears in the 5D action. 
A detailed derivation of the effective K\"{a}hler potential 
is provided in Appendix~\ref{eff_Kahler}.

\section{4D effective theory with multi moduli} 
\label{derive_4DEFT} 
In this section we briefly review the off-shell 
dimensional reduction~\cite{Abe:2006eg} 
and derive the 4D effective action of 
5D SUGRA compactified on an orbifold~$S^1/Z_2$ 
with an arbitrary norm function. 
The 5D metric is assumed to be 
\be
 ds^2 = e^{2\sgm(y)}g_{\mu\nu}dx^\mu dx^\nu-\brkt{\gey dy}^2, 
\ee
where $\mu,\nu=0,1,2,3$, and $e^{\sgm(y)}$ is a warp factor, 
which is a function of only $y$ 
and determined by the dynamics. 
We take the fundamental region of the orbifold as $0\leq y\leq\pi R$, 
where $R$ is a constant.\footnote{
In principle, $R$ is nothing to do with the radius of the orbifold~$r$, 
which is given by the proper length along the fifth coordinate. 
}

\subsection{$N=1$ off-shell description of 5D SUGRA action}
Our formalism is based on the 5D conformal SUGRA formulation 
in Ref.~\cite{Kugo-Ohashi,Kugo:2002js}. 
5D superconformal multiplets relevant to our study are 
the Weyl multiplet~$\bdm{E}_W$, vector multiplets~$\bdm{V}^I$, 
and hypermultiplets~$\bdm{H}^{\hat{a}}$, 
where $I=1,2,\cdots,n_V$ and $a=1,2,\cdots,n_C+n_H$. 
Here $n_C$ and $n_H$ are the numbers of compensator and 
physical hypermultiplets, respectively. 
These 5D multiplets are decomposed into $N=1$ superconformal 
multiplets~\cite{Kugo:2002js} as $\bdm{E}_W=(E_W,L^\alp,V_E)$, 
$\bdm{V}^I=(V^I,\Sigma^I)$ and 
$\bdm{H}^a=(\Phi^{2a-1},\Phi^{2a})$, 
where $E_W$ is the $N=1$ Weyl multiplet, $V_E$ is an $N=1$ real general 
multiplet whose scalar component is $\gey$, 
$V^I$ is an $N=1$ vector multiplet, and $\Sigma^I$, 
$\Phi^{\hat{a}}$ ($\hat{a}=1,2,\cdots,2(n_C+n_H)$) 
are chiral multiplets. 
An $N=1$ complex general multiplet~$L^\alp$ ($\alp$: spinor index) 
in $\bdm{E}_W$ consists of the $Z_2$-odd components, 
and is irrelevant to the following discussion 
(see Appendix~\ref{Z2odd_mul}). 
Thus we neglect its dependence of the 5D action in the following. 

The 5D SUGRA action can be written in terms of these $N=1$ 
multiplets~\cite{Paccetti Correia:2004ri,Abe:2004}. 
Then we can see that $V_E$ has no kinetic term.\footnote{
This does not mean that $\gey$ is an auxiliary field. 
It is also contained in $\Sgm^I$ ($I=1,2,\cdots,n_V$), 
which have their own kinetic terms. } 
After integrating $V_E$ out, the 5D action is expressed 
as~\cite{Correia:2006pj} 
\begin{eqnarray}
 \cL \eql 
 -\frac{1}{4}\sbk{\int\dr^2\tht\;\cN_{IJ}(\Sgm)\cW^I\cW^J+\hc}
 +\cdots \nonumber\\
 &&-3e^{2\sigma} \int\dr^4\tht\,\cN^{1/3}(\cV) 
 \brc{d_{\hat{a}}^{\;\; \hat{b}} \bar{\Phi}^{\hat{b}} 
 \brkt{e^{-2igt_IV^I}}^{\hat{a}}_{\;\; \hat{c}} \Phi^{\hat{c}}}^{2/3} 
 \nonumber \\ 
 &&-e^{3\sigma} \bigg[ 
 \int\dr^2 \theta\, \Phi^{\hat{a}}d_{\hat{a}}^{\;\; \hat{b}} 
 \rho_{\hat{b}\hat{c}} 
 \brkt{\partial_y-2igt_I\Sigma^I}^{\hat{c}}_{\;\; \hat{d}} \Phi^{\hat{d}} 
 +\textrm{h.c.} \bigg] \nonumber\\
 &&+\sum_{\vtht=0,\pi}\cL_{\vtht}\,\dlt(y-\vtht R), 
\label{eq:lbulk}
\end{eqnarray}
where 
$d_{\hat{a}}^{\;\; \hat{b}}={\rm diag}({\bf 1}_{2n_C},-{\bf 1}_{2n_H})$  
and $\rho_{\hat{a}\hat{b}}=i \sigma_2 \otimes {\bf 1}_{n_C+n_H}$. 
Here $\sgm_2$ acts on each hypermultiplet~$(\Phi^{2a-1},\Phi^{2a})$. 
$\cN$ is a cubic polynomial called the norm function, 
which is defined by 
\be
 \cN(X) \equiv C_{IJK}X^IX^JX^K.  \label{def_cN}
\ee
A real constant tensor~$C_{IJK}$ is completely symmetric for the indices, 
and $\cN_{IJ}(X)\equiv\der^2\cN/\der X^I\der X^J$. 
The superfield strength~$\cW^I$ and 
$\cV^I\equiv -\der_y V^I+\Sgm^I+\bar{\Sgm}^I$ 
are gauge-invariant quantities. 
The generators~$t_I$ are anti-hermitian. 
The ellipsis in (\ref{eq:lbulk}) denotes the supersymmetric 
Chern-Simons terms that are irrelevant to the following discussion. 
The boundary Lagrangian~$\cL_\vtht$ ($\vtht=0,\pi$) can be introduced 
independently of the bulk action. 
Note that (\ref{eq:lbulk}) is a shorthand expression for 
the full SUGRA action. 
We can always restore the full action by promoting 
the $d^4\tht$ and $d^2\tht$ 
integrals to the $D$- and $F$-term action formulae 
of the $N=1$ conformal SUGRA formulation~\cite{4Doffshell}, 
which are compactly listed in Appendix~C of Ref.~\cite{Kugo:2002js}. 

The vector multiplets~$\bdm{V}^I$ are classified into 
$(\bdm{V}^{I'},\bdm{V}^{I''})$ by their orbifold parities so that 
$V^{I'}$ ($I'=1,2,\cdots,n_V'$) and $V^{I''}$ ($I''=n_V'+1,\cdots,n_V$) 
are odd and even, respectively.  
As for the hypermultiplets~$(\Phi^{2a-1},\Phi^{2a})$, 
we can always choose the orbifold parities as listed 
in Table~\ref{Z2_parity} by using $\SUu$, which is an automorphism 
in the superconformal algebra. 
\begin{table}[t]
\begin{center}
\begin{tabular}{|c|c|c|c||c|c|} \hline
\rule[-2mm]{0mm}{7mm} $V^{I'}$ & $\Sgm^{I'}$ & $V^{I''}$ & $\Sgm^{I''}$ 
& $\Phi^{2a-1}$ & $\Phi^{2a}$ \\ \hline
$-$ & $+$ & $+$ & $-$ & $-$ & $+$ \\ \hline
\end{tabular}
\end{center}
\caption{The orbifold parities for $N=1$ multiplets. }
\label{Z2_parity}
\end{table}
As explained in Ref.~\cite{Abe:2006eg}, $n_V'$ moduli come out 
from $\Sgm^{I'}$ in the 4D effective theory. 
In the case of $n_V'=1$, the corresponding modulus is identified with 
the radion multiplet. 
The 4D massless gauge fields, such as the standard model gauge fields, 
come out from $V^{I''}$. 
For a gauge multiplet of a nonabelian gauge group~$G$, 
the indices~$I''$ and $J''$ run over ${\rm \dim}\, G$ values and 
$\cN_{I''J''}$ are common for them. 
The index~$a$ for the hypermultiplets are divided into 
irreducible representations of $G$. 
In the following we consider a case that $n_V'=2$ and $n_C=1$ 
as the simplest case with multi moduli. 
An extension to the cases that $n_V'>2$ is straightforward. 

In 5D SUGRA, every mass scale in the bulk action is introduced 
by gauging some of the isometries on the hyperscalar manifold\footnote{ 
The hyperscalar manifold is $USp(2,2n_H)/USp(2)\times USp(2n_H)$ 
for $n_C=1$, and $SU(2,1)/SU(2)\times U(1)$ for $n_C=2$. 
}
by some vector multiplets~$\bdm{V}^{I'}$.  
For example, the bulk cosmological constant is induced 
when the compensator multiplet~$(\Phi^1,\Phi^2)$ is charged, 
and a bulk mass parameter for a physical hypermultiplet is induced 
when it is charged for $\bdm{V}^{I'}$. 
Of course, we can also gauge some of the isometries by $\bdm{V}^{I''}$. 
This leads to the usual gauging for the chiral multiplets 
by a 4D massless gauge multiplet in the 4D effective theory. 
In the following we will omit the $\bdm{V}^{I''}$-dependence of the action 
except for the kinetic terms 
because it does not play a significant role in the procedure of 
the off-shell dimensional reduction 
and can be easily restored in the 4D effective action. 
In this paper we consider a case that 
only the physical hypermultiplets~$(\Phi^{2a-1},\Phi^{2a})$ 
($a=2,\cdots,n_H+1$) are charged for $\bdm{V}^{I'}$ ($I'=1,2$). 
This corresponds to a flat background geometry of 5D spacetime. 
We assume that all the directions of the gauging are chosen 
to $\sgm_3$-direction for $(\Phi^{2a-1},\Phi^{2a})$ 
since the gauging along the other directions 
mixes $\Phi^{2a-1}$ and $\Phi^{2a}$, which have opposite parities. 
Namely, the generators and the gauge couplings are chosen as 
\be
 \brkt{igt_{I'}}^{\hat{a}}_{\;\;\hat{b}} 
 = \sgm_3\otimes \diag(0,c_{2I'},c_{3I'},\cdots,c_{(n_H+1)I'}), 
\ee
where $\sgm_3$ acts on each hypermultiplet~$(\Phi^{2a-1},\Phi^{2a})$, 
and $c_{aI'}$ ($a=2,\cdots,n_H+1$) are gauge coupling constants 
for $\bdm{V}^{I'}$. 

Then, after rescaling chiral multiplets by a factor $e^{3\sgm/2}$, 
we obtain 
\bea
 \cL \eql -\frac{1}{4}\sbk{\int\dr^2\tht\:\cN_{IJ}(\Sgm)\cW^I\cW^J+\hc}
 +\cdots \nonumber\\
 &&-3\int\dr^4\tht\;\cN^{1/3}(\cV)
 \left\{\abs{\Phi^1}^2+\abs{\Phi^2}^2 \right. \nonumber\\
 &&\hspace{30mm}\left.-\sum_{a=2}^{n_H+1}
 \brkt{\bar{\Phi}^{2a-1}e^{-2c_a\cdot V}\Phi^{2a-1}
 +\bar{\Phi}^{2a}e^{2c_a\cdot V}\Phi^{2a}}\right\}^{2/3} \nonumber\\
 &&-2\left[\int\dr^2\tht\;\brc{
 \Phi^1\der_y\Phi^2 
 -\sum_{a=2}^{n_H+1}\Phi^{2a-1}\brkt{
 \der_y+2c_a\cdot \Sgm}\Phi^{2a}}+\hc \right]
 \nonumber\\
 &&+\sum_{\vtht=0,\pi}\sbk{\int\dr^2\tht\;
 \brkt{\Phi^2}^2 W^{(\vtht)}(\hat{\Phi}^{2a})+\hc}, 
\eea
where 
$c_a\cdot V\equiv \sum_{I'=1}^2c_{aI'}V^{I'}$, and 
$\hat{\Phi}^{2a}\equiv \Phi^{2a}/\Phi^2$.  
For simplicity, we have introduced 
only superpotentials~$W^{(\vtht)}$ in the boundary Lagrangians. 
The hypermultiplets appear in $W^{(\vtht)}$ 
only through $\hat{\Phi}^{2a}$ because 
physical chiral multiplets must have zero Weyl weights 
in the $N=1$ conformal SUGRA~\cite{4Doffshell} 
and $\Phi^{2a-1}$ vanish at the boundaries due to their orbifold parities.

\subsection{4D effective action}
Following the procedure explained in Sec.~3 of Ref.~\cite{Abe:2006eg}, 
we can derive the 4D effective action. 
First, we remove all $\Sgm^I$ from the bulk action 
by 5D gauge transformation. 
Since $\Sgm^{I'}$ are even under the orbifold parity and thus 
have zero-modes, the gauge transformation parameters~$\Lmd^{I'}$ 
must be discontinuous at one of the boundaries. 
In the notation of Ref.~\cite{Abe:2006eg}, 
$\Lmd^{I'}$ are discontinuous at $y=\pi R$. 
The gaps correspond to zero-modes for $\Sgm^{I'}$. 
Such zero-modes are called the moduli~$T^{I'}$ in this paper, 
and defined by\footnote{
The normalization of $T^{I'}$ is different from that 
in Ref.~\cite{Abe:2006eg} by a factor~$\pi$. }  
\be
 T^{I'} \equiv 2\int_0^{\pi R}\dr y\;\Sgm^{I'}(y). 
\ee 
Namely, 
\be
 \lim_{\ep\to +0}\Lmd^{I'}(y=\pi R-\ep) = -\frac{1}{2}T^{I'}, 
\ee
which means 
\be
 \lim_{\ep\to +0}\tl{V}^{I'}(y=\pi R-\ep) = -\Re T^{I'}, 
 \label{BC_V:pi}
\ee
where $\tl{V}^{I'}$ are the $N=1$ vector multiplets 
after the gauge transformation. 
Since $\Lmd^{I'}$ are continuous at the other boundary ($y=0$), 
$\tl{V}^{I'}$ obey ordinary Dirichlet boundary conditions there. 
\be
 \tl{V}^{I'}(y=0) = 0. \label{BC_V:0}
\ee

Next we neglect the kinetic terms for parity-odd $N=1$ multiplets 
because they do not have zero-modes which are dynamical below 
the compactification scale. 
Then the parity-odd multiplets play a role of Lagrange multipliers 
and their equations of motion extract zero-modes from 
the parity-even multiplets.\footnote{
The effects of the parity-odd multiplet~$L^\alp$ in the 5D Weyl multiplet 
on the effective theory are negligible because it couples to the matter 
multiplets only in the derivative forms (see Appendix~\ref{Z2odd_mul}). 
} 
In fact, $\tl{V}^{I''}$, $\Phi^2$ and $\hat{\Phi}^{2a}$ become 
$y$-independent and are identified with 4D zero-modes after 
the parity-odd fields are integrated out.  
By performing the $y$-integral, the following expression is obtained.  
\bea
 \cL^{(4D)} \eql -\frac{3}{4}\sbk{\int\dr^2\tht\;
 C_{I'J''K''}T^{I'}\tl{\cW}^{J''}\tl{\cW}^{K''}+\hc} \nonumber\\
 &&-3\int\dr^4\tht\;\abs{\phi}^2
 \brc{\int_0^{\pi R-\ep}\dr y\;\hat{\cN}^{1/3}(-\der_y\tl{V})
 \brkt{1-\sum_{a=2}^{n_H+1}e^{2c_a\cdot\tl{V}}\abs{\hat{\Phi}^{2a}}^2}^{2/3}} 
 \nonumber\\
 &&+\sbk{\int\dr^2\tht\;\phi^3
 \brc{W^{(0)}(\hat{\Phi})+W^{(\pi)}(e^{-c_a\cdot T}\hat{\Phi})}+\hc}, 
 \label{L_4D:1}
\eea
where $\phi\equiv (\Phi^2)^{2/3}$ is the 4D chiral compensator multiplet, 
and $\hat{\cN}$ is a truncated function of the norm function 
defined by
\be
 \hat{\cN}(X) \equiv C_{I'J'K'}X^{I'}X^{J'}X^{K'}.
\ee
Notice that $\tl{V}^{I'}$ ($I'=1,2$) still have $y$-dependence while 
the other fields are now $y$-independent. 
In the single modulus case (\ie, $n_V'=1$), the $y$-integral in 
(\ref{L_4D:1}) can be easily performed because the integrand becomes 
a total derivative for $y$~\cite{Correia:2006pj}. 
On the other hand, in the multi moduli case (\ie, $n_V'\geq 2$), 
$\tl{V}^{I'}$ must be integrated out 
by using their equations of motion~\cite{Abe:2006eg}. 
To simplify the discussion, let us assume that each hypermultiplet 
is charged for only one of $\bdm{V}^{I'}$ ($I'=1,2$). 
Namely we can classify the physical hypermultiplets~$\hat{\Phi}^{2a}$ 
($a=2,\cdots,n_H+1$) 
into $Q_i$ ($i=1,2,\cdots,n_1$) and $S_\alp$ ($\alp=1,2,\cdots,n_2$) 
so that $Q_i$ are charged for $\bdm{V}^1$ and $S_\alp$ are charged 
for $\bdm{V}^2$. 
Here $n_H=n_1+n_2$. 
Then the parenthesis in the second line of (\ref{L_4D:1}) is rewritten as 
\be
 \brkt{1-\sum_{i=1}^{n_1}e^{2c_i\tl{V}^1}\abs{Q_i}^2
 -\sum_{\alp=1}^{n_2}e^{2c_\alp\tl{V}^2}\abs{S_\alp}^2}^{2/3}. 
 \label{brkt:QS}
\ee
The detailed calculations are summarized in Appendix~\ref{eff_Kahler}. 
The result is 
\bea
 \cL^{(4D)} \eql -\frac{1}{4}\sbk{\int\dr^2\tht\;
 \sum_rf_r(T)\tr\brkt{\cW^r\cW^r}+\hc} \nonumber\\
 &&+\int\dr^4\tht\;\abs{\phi}^2\Omg(\abs{Q}^2,\abs{S}^2,\Re T)
 +\sbk{\int\dr^2\tht\;\phi^3W(Q,S,T)+\hc}, 
 \label{eff_cL}
\eea
where the vector multiplets are summarized in the matrix forms 
for the nonabelian gauge multiplets and the index~$r$ indicates 
different gauge multiplets. 
Each function in (\ref{eff_cL}) is defined as 
\bea
 f_r \defa \sum_{I'=1}^2k_{rI'}T^{I'}, \nonumber\\
 \Omg \defa -3\hat{\cN}^{1/3}(\Re T)
 \brc{1-\sum_i\frac{1-e^{-2c_i\Re T^1}}{3c_i\Re T^1}\abs{Q_i}^2
 -\sum_\alp\frac{1-e^{-2c_\alp\Re T^2}}{3c_\alp\Re T^2}\abs{S_\alp}^2}
 \nonumber\\
 &&+\sum_{i,j}\Omg_{ij}\abs{Q_i}^2\abs{Q_j}^2
 +\sum_{i,\alp}\Omg_{i\alp}\abs{Q_i}^2\abs{S_\alp}^2
 +\sum_{\alp,\bt}\Omg_{\alp\bt}
 \abs{S_\alp}^2\abs{S_\bt}^2+\cO(\hat{\Phi}^6),  \nonumber\\
 W \defa W^{(0)}(\hat{\Phi})+W^{(\pi)}(e^{-c_a\cdot T}\hat{\Phi}), 
 \label{eff_def_fcn}
\eea
where $k_{rI'}$ are constants determined from $C_{I'J''K''}$, 
$\hat{\Phi}=Q_i$ or $S_\alp$, and 
\bea
 \Omg_{ij} \defa 
 \frac{\hat{\cN}^{4/3}\hat{\cN}_{11}}
 {3\hat{\cN}\hat{\cN}_{11}-2\hat{\cN}_1^2}
 \frac{1-e^{-2(c_i+c_j)\Re T^1}}{2(c_i+c_j)\Re T^1} \nonumber\\
 &&-\frac{\hat{\cN}^{1/3}\hat{\cN}_1^2}
 {3\hat{\cN}\hat{\cN}_{11}-2\hat{\cN}_1^2}
 \frac{(1-e^{-2c_i\Re T^1})(1-e^{-2c_j\Re T^1})}
 {6c_ic_j(\Re T^1)^2}, \nonumber\\
 \Omg_{i\alp} \defa 
 \frac{\hat{\cN}^{4/3}\hat{\cN}_{12}}
 {3\hat{\cN}\hat{\cN}_{12}-2\hat{\cN}_1\hat{\cN}_2}
 \frac{1-e^{-2c_i\Re T^1-2c_\alp\Re T^2}}{c_i\Re T^1+c_\alp\Re T^2} 
 \nonumber\\
 &&-\frac{\hat{\cN}^{1/3}\hat{\cN}_1\hat{\cN}_2}
 {3\hat{\cN}\hat{\cN}_{12}-2\hat{\cN}_1\hat{\cN}_2}
 \frac{(1-e^{-2c_i\Re T^1})(1-e^{-2c_\alp\Re T^2})}
 {3c_ic_\alp\Re T^1\Re T^2}, \nonumber\\
 \Omg_{\alp\bt} \defa 
 \frac{\hat{\cN}^{4/3}\hat{\cN}_{22}}
 {3\hat{\cN}\hat{\cN}_{22}-2\hat{\cN}_2^2}
 \frac{1-e^{-2(c_\alp+c_\bt)\Re T^2}}{2(c_\alp+c_\bt)\Re T^2} \nonumber\\
 &&-\frac{\hat{\cN}^{1/3}\hat{\cN}_2^2}
 {3\hat{\cN}\hat{\cN}_{22}-2\hat{\cN}_2^2}
 \frac{(1-e^{-2c_\alp\Re T^2})(1-e^{-2c_\bt\Re T^2})}
 {6c_\alp c_\bt(\Re T^2)^2}. 
 \label{cf_Omg}
\eea
Here $\hat{\cN}_{I'}(X)\equiv d\hat{\cN}/dX^{I'}$ 
and $\hat{\cN}_{I'J'}(X)\equiv d^2\hat{\cN}/dX^{I'}dX^{J'}$. 
The arguments of $\hat{\cN}$, $\hat{\cN}_{I'}$, $\hat{\cN}_{I'J'}$ 
are understood as $(\Re T^1,\Re T^2)$. 

When all the gauge couplings~$c_i,c_\alp$ vanish, the exact form 
of $\Omg$ is obtained as (\ref{simple_Omg}). 
We can easily check that 
the above result is consistent with (\ref{simple_Omg}) 
by taking the limit of $c_i,c_\alp\to 0$. 

Some of the nonabelian gauge multiplets may condense 
and generate superpotential terms at low energies, 
which have a form of $\exp\{-a_rf_r(T)\}$ where $a_r=\cO(4\pi^2)$, 
because the gauge kinetic function~$f_r$ is proportional to 
inverse square of the gauge coupling. 
(See eq.(\ref{W_np}).) 

Before ending this section, we note that there is no 
``radion chiral multiplet'' in the multi moduli case. 
In the 5D conformal SUGRA, we have to fix the extra symmetries 
by imposing gauge-fixing conditions in order to obtain the Poincar\'{e} 
supergravity.\footnote{
In the procedure of the off-shell dimensional reduction, 
we do not impose such gauge-fixing conditions 
to keep the $N=1$ off-shell structure. 
They should be imposed {\it after} the 4D effective action is obtained. } 
The dilatation symmetry is fixed by a condition: 
\be
 \cN(M)=C_{IJK}M^IM^JM^K=1, 
\ee
in the unit of the 5D Planck mass, 
where $M^I$ is a real scalar component of $\bdm{V}^I$ 
and is related to a scalar component of $\Sgm^I$ by 
$M^I\equiv 2\Re\Sgm^I|_{\theta=0}/\gey$. 
This means that the size of the orbifold~$\pi r$ is determined by 
\be
 \pi r \equiv \int_0^{\pi R}\dr y\;\gey
 = 
 \left.2\int_0^{\pi R}\dr y\;\cN^{1/3}(\Re\Sgm)\right|_{\tht=0}
 \simeq \left.\hat{\cN}^{1/3}(\Re T)\right|_{\tht=0}. 
\ee
The last equation holds when the background geometry of 5D spacetime 
is flat and the backreaction to the geometry due to the 5D scalar field 
configurations is negligible. 
In the single modulus case (\ie, $\cN=\hat{\cN}=(M^1)^3$), 
the above relation is reduced to 
\be
 \pi r = \Re T^1|_{\tht=0}, 
\ee
which means that $T^1$ is the radion multiplet. 
In the multi moduli case, on the other hand, we cannot redefine 
a chiral multiplet whose scalar component gives the size of the orbifold
by holomorphic redefinition. 
It is given by a combination of VEVs of all the moduli. 
In other words, the radion mode cannot form an $N=1$ chiral multiplet 
without mixing with the other moduli in the multi moduli case.

\section{Soft SUSY breaking terms} \label{soft_terms}
\subsection{Flavor structure of soft parameters} \label{flavor_soft}
In this section we discuss the flavor structure of 
the soft SUSY breaking terms. 
We introduce a chiral multiplet~$X$ that is relevant to SUSY breaking, 
in addition to the MSSM field content which consists of  
the gauge multiplets~$V^r$ ($r=1,2,3$) and 
the matter chiral multiplets~$\cS_l$, where $l$ runs over quark, lepton 
and Higgs multiplets. 
Each chiral multiplet~$\cS_l$ can be either $Q_i$ or $S_\alp$ 
in (\ref{eff_def_fcn}). 
We identify $X$ as one of $Q_i$ without loss of generality. 
We take the unit of the 4D Planck mass, \ie, $M_{\rm Pl}=1$, 
in the rest of this paper. 

The Yukawa couplings among $\cS_l$ can be introduced only 
at the orbifold boundaries due to the $N=2$ SUSY in the bulk. 
Here we assume that they exist only at one boundary ($y=0$), 
for simplicity.  
Namely, we introduce the following boundary superpotential. 
\be
 W^{(0)}_{\rm yukawa} = \sum_{a,b,c}\lmd_{abc}
 \hat{\Phi}^{2a}\hat{\Phi}^{2b}\hat{\Phi}^{2c}, 
 \label{bd_Yukawa}
\ee
where $\lmd_{abc}$ are constants.\footnote{
In general, $\lmd_{abc}$ can depend on $X$, but we do not consider 
this possibility, for simplicity. } 

Here we focus on the gaugino masses~$M_r$ ($r=1,2,3$), 
the scalar masses~$m_l$ and the A-parameters~$A_{lmn}$, 
which are defined as 
\be
 \cL_{\rm soft} = -\sum_l m^2_l \abs{\tl{\cS}_l}^2
 -\frac{1}{2}\brc{\sum_rM_r\tl{\lmd}^r\tl{\lmd}^r
 +\frac{1}{6}\sum_{l,m,n}y_{lmn}A_{lmn}\tl{\cS}_l\tl{\cS}_m\tl{\cS}_n+\hc}, 
 \label{def:soft_pt}
\ee
where $\tl{\cS}_l$, $\tl{\lmd}^r$ are {\it canonically normalized} 
sfermions and gauginos, and $y_{lmn}$ are the physical Yukawa coupling 
constants for the canonically normalized fields. 
These soft SUSY breaking terms are generated through the mediation by 
the moduli~$T^{I'}$ ($I'=1,2$) as well as the direct couplings 
to the SUSY breaking superfield~$X$. 

Let us rewrite $\Omg$ in (\ref{eff_def_fcn}) as 
\bea
 \Omg \eql \Omg_0(\Re T)+\sum_lY_l(\Re T,\abs{X}^2)\abs{\cS_l}^2
 +\cO(\cS^4), 
\eea
where 
\bea
 \Omg_0(\Re T) \defa -3\hat{\cN}^{1/3}(\Re T), \nonumber\\
 Y_l(\Re T,\abs{X}^2) \defa \hat{\cN}^{1/3}(\Re T)Z_l(\Re T^1)
 +\Omg_{lX}(\Re T)\abs{X}^2+\cO(\abs{X}^4), 
 \label{Omg_expand}
\eea
with 
\be
 Z_l(\Re T) \equiv \begin{cases} 
 {\displaystyle \frac{1-e^{-2c_l\Re T^1}}{c_l\Re T^1}}, & 
 \mbox{when $\cS_l\in\{Q_i\}$} \\*[5pt]
 {\displaystyle \frac{1-e^{-2c_l\Re T^2}}{c_l\Re T^2}}, & \mbox{when $\cS_l\in\{S_\alp\}$} 
 \end{cases}
\ee
and 
\be
 \Omg_{lX} = \begin{cases} \Omg_{i=X,j=l}+\Omg_{i=l,j=X}=2\Omg_{i=X,j=l} 
 & \mbox{when $\cS_l\in\{Q_i\}$}  \\ 
 \Omg_{i=X,\alp=l} & \mbox{when $\cS_l\in\{S_\alp\}$} \end{cases}
\ee
Then the physical Yukawa couplings and 
the soft parameters in (\ref{def:soft_pt}) are
expressed in terms of $Y_l(\Re T,\abs{X}^2)$ as~\cite{CFNO,KL} 
\bea
 y_{lmn} \defa \frac{\lmd_{lmn}}{\sqrt{Y_lY_mY_n}}, \nonumber\\
 M_r \defa F^A\der_A\ln\brkt{\Re f_r}, \nonumber\\
 m_l^2 \defa -F^A\bar{F}^{\bar{B}}\der_A\der_{\bar{B}}\ln Y_l, \nonumber\\
 A_{lmn} \defa -F^A\der_A\ln\brkt{Y_lY_mY_n}, 
 \label{fml_softSB}
\eea
where indices~$A,B$ run over all the chiral multiplets. 
The hierarchical structure of the Yukawa couplings is realized 
by varying $c_l$ in an $\cO(1)$ range (see, e.g., \cite{CKKK,ACJO}). 
The small fermion masses for $\cS_l$ are obtained 
by taking $c_l$ negative so that 
$Z_l$ are large enough.\footnote{
In general, $\Re T^{I'}$ can be negative as long as $\cN(\Re T)$ is positive. 
However we assume that $\Re T^{I'}>0$ ($I'=1,2$) in the following. } 
For the Higgs multiplets, $c_l$ is taken to be positive 
in order to realize the large top quark mass. 

Let us assume that $F^X$ is a dominant source of SUSY breaking, \ie, 
\be
 \abs{F^X}\gg \abs{F^{T^1}}, \abs{F^{T^2}}. \label{F:assumption}
\ee
This is indeed the case in the model considered in Sec.~\ref{moduli_stb}. 
Then the soft scalar masses are given by 
\bea
 m_l^2 \sima -\abs{F^X}^2\frac{\der_X\der_{\bar{X}}Y_l}{Y_l}
 \simeq -\abs{F^X}^2\frac{\Omg_{lX}}{\hat{\cN}^{1/3}Z_l}. 
 \label{ap_ml}
\eea
We have assumed that $\abs{X}\ll 1$ in order for the expansion of 
$Y_l$ in (\ref{Omg_expand}) to be valid,  
which is also realized in the model in Sec.~\ref{moduli_stb}. 

In the single modulus case, $\Omg_{lX}$ is calculated as 
\be
 \Omg_{lX} = \hat{\cN}^{1/3}(\Re T^1)\frac{1-e^{-2(c_l+c_X)\Re T^1}}
 {3(c_l+c_X)\Re T^1} 
 = \frac{1-e^{-2(c_l+c_X)\Re T^1}}{3(c_l+c_X)}. 
 \label{Omg_single}
\ee
Notice that this is always positive\footnote{
Since $T^1$ is the radion in this case, $\Re T^1$ must be stabilized 
at a positive value. } 
irrespective of the values of $c_l$ and $c_X$. 
This means that the soft scalar masses are tachyonic~\cite{single_modulus}. 
This tree-level contribution~(\ref{ap_ml}) is exponentially suppressed 
when $e^{-2c_l\Re T^1}\gg 1$ and $e^{-2c_X\Re T^1}\ll 1$, 
which corresponds to the case that the visible matter multiplets~$\cS_l$ 
and the SUSY breaking multiplet~$X$ are localized around the opposite 
boundaries. 
In such a case, quantum effects to the soft scalar masses become dominant 
and may save the tachyonic masses at tree-level. 
However the large top quark mass cannot be realized in this case 
because the top Yukawa coupling is suppressed by the large $Y_l$. 

This problem can be evaded in the multi moduli case 
because $\Omg_{lX}$ is modified. 
We explain the situation by two examples of the norm function. 
In the following we assume that $c_X>0$, \ie, $e^{-2c_X\Re T^{I'}}\ll 1$. 
\begin{description}
 \item{\bf Example 1:}
\be
 \hat{\cN}(X) = (X^1)^3+(X^2)^3 \label{norm1}
\ee
In the case that $\cS_l\in\brc{Q_i}$, 
the first term of $\Omg_{lX}$ is positive 
while the second term negative 
since $3\hat{\cN}\hat{\cN}_{11}-2\hat{\cN}_1^2>0$. 
For the first and the second generations, $e^{-2c_l\Re T^1}\gg 1$ 
to realize the small fermion masses. 
Then the second term dominates and the soft squared masses become positive. 
Furthermore the soft masses are almost degenerate because 
the $c_l$-dependence are cancelled in (\ref{ap_ml}) 
when the second term of $\Omg_{lX}$ dominates. 
For the top quark, on the other hand, 
we have to take $c_l$ such that $e^{-2c_l\Re T^1}\simlt\cO(1)$ 
in order to realize the large top quark mass. 
Thus $\Omg_{lX}$ is positive, which leads to  
the tachyonic stop masses. 

In the case that $\cS_l\in\brc{S_\alp}$, 
the sign of $\Omg_{lX}$ becomes opposite because of 
the identity~(\ref{identity2}). 
Therefore the stop masses are now nontachyonic. 

In summary, if we take $\cS_l\in\brc{S_\alp}$ for the top quark multiplets 
and $\cS_l\in\brc{Q_i}$ for the other multiplets, 
all the soft masses are nontachyonic and 
they are almost degenerate for the first two generations. 
Since the severest constraints on the soft masses come from 
the flavor changing processes for the first two generations, 
this setup can solve the SUSY flavor problem. 

 \item{\bf Example 2:}
\be
 \hat{\cN}(X) = (X^1)^2X^2 \label{norm2}
\ee
In this case, the situations for $\cS_l\in\brc{Q_i}$ 
and for $\cS_l\in\brc{S_\alp}$ in Example~1 are interchanged 
since now $3\hat{\cN}\hat{\cN}_{11}-2\hat{\cN}_1^2<0$. 
In summary, we can construct a phenomenologically viable model 
if we take $\cS_l\in\brc{Q_i}$ for the top quark multiplets 
and $\cS_l\in\brc{S_\alp}$ for the other multiplets.  
\end{description}
We comment on the possibility to choose $\hat{\cN}$ 
such that $\Omg_{lX}$ is negative for any values of $c_l$ and $c_X$. 
From (\ref{cf_Omg}), such $\hat{\cN}$ must satisfy 
$\hat{\cN}_{11}<0$ and $3\hat{\cN}\hat{\cN}_{11}-2\hat{\cN}_1^2>0$ 
in the case that $\cS_l\in\brc{Q_i}$, 
or $\hat{\cN}_{12}/(3\hat{\cN}\hat{\cN}_{12}-2\hat{\cN}_1\hat{\cN}_2)<0$ 
and $\hat{\cN}_1\hat{\cN}_2/
(3\hat{\cN}\hat{\cN}_{12}-2\hat{\cN}_1\hat{\cN}_2)>0$ 
in the case that $\cS_l\in\brc{S_\alp}$. 
However the two conditions are incompatible in either case. 
Therefore the the first two generations and the top quark multiplets 
must be charged for different gauge multiplets~$\bdm{V}^{I'}$ 
in order to avoid the tachyonic soft masses. 

As for the A parameters, contribution from $F^X$ is negligible 
because it accompanies with VEV of $\bar{X}$, which is assumed to be tiny. 
Thus the dominant contributions come from $F^{T^{I'}}$ ($I'=1,2$) and 
are, in general, flavor dependent. 
The resultant A-parameters are much smaller than 
the soft masses~$m_l$ due to the assumption~(\ref{F:assumption}). 
Furthermore, the flavor dependence of the A-parameters 
can be small if there is a hierarchy between $F^{T^1}$ and $F^{T^2}$. 
For instance, let us assume that $|F^{T^1}|\gg |F^{T^2}|$ 
and $Y_l$ are almost independent of the first modulus~$T^1$. 
These conditions are satisfied for the first two generations 
in Example~2. 
(See eq.(\ref{Fs:eg2}).)
Then the A-parameters are estimated as 
\bea
 A_{lmn} \sima -F^{T^1}\brkt{\frac{\der_{T^1}Y_l}{Y_l}
 +\frac{\der_{T^1}Y_m}{Y_m}+\frac{\der_{T^1}Y_n}{Y_n}} \nonumber\\
 \sima -3F^{T^1}\der_{T^1}\brc{\hat{\cN}^{1/3}(\Re T)} 
 = -F^{T^1}\frac{\hat{\cN}_1}{2\hat{\cN}^{2/3}}(\Re T),  
\eea
which are almost independent of the flavor indices. 

The gaugino masses are estimated to be the same order as 
the A-parameters because the gauge kinetic functions~$f_r$ only depend 
on the moduli~$T^{I'}$. 
The situation can be changed by introducing the gauge kinetic functions 
that depend on $X$ at the boundaries. 

The soft SUSY breaking parameters obtained by (\ref{fml_softSB}) 
should be understood as those at the compactification scale, 
which is close to the Planck scale~$M_{\rm Pl}$ 
when $\Re T^{I'}=\cO(1)$. 
Thus we have to evaluate the soft parameters at $M_{\rm weak}$ 
obtained by RGEs 
in order to check whether the tachyonic sfermion mass problem and 
the SUSY flavor problem are really solved or not. 
We will discuss this issue by numerical calculations 
in a specific model in Sec.~\ref{num_calc}.

\subsection{Interpretation of $\bdm{\Omg_{lX}}$ from 5D viewpoint}
The essential difference between the single and multi moduli cases 
appears in the form of $\Omg_{lX}$.
Especially the second term of $\Omg_{lX}$ in the multi moduli case 
has a peculiar flavor structure. 
Here we give an interpretation of it from the 5D viewpoint. 

The condition under which the second term of $\Omg_{lX}$ dominates, \ie, 
$e^{-2c_l\Re T^{I'}}\gg 1$ and $e^{-2c_X\Re T^{I'}}\ll 1$, 
corresponds to a situation in which the zero-modes~$\cS_l$ 
and $X$ are geometrically separated from each other. 
In fact, in such a situation, 
the contact interactions~$\Omg_{lX}$ are exponentially 
suppressed in the single modulus case~\cite{single_modulus}. 
This suggests that in the multi moduli case, 
there exist some heavy modes that couple to both $\cS_l$ and $X$, 
which induce contact interactions after they are integrated out. 
Such heavy modes are identified with 
the parity-odd vector multiplets~$V^{I'}$. 

\begin{figure}[t,b]
\centering \leavevmode
\includegraphics[width=70mm]{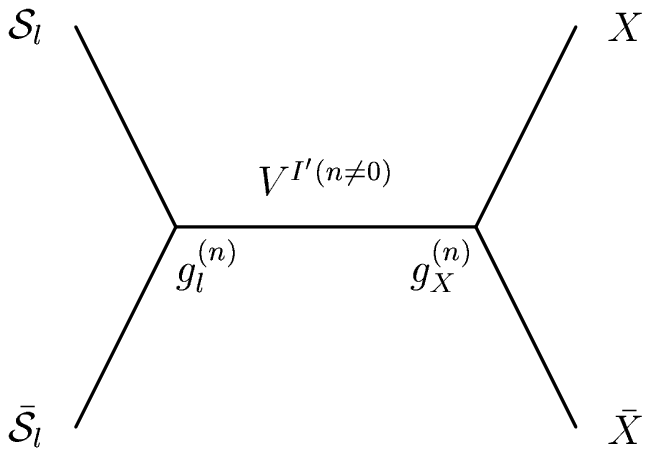}
\caption{Feynmann diagrams contributing to $\Omg_{lX}$ 
in the multi moduli case. The index~$n$ labels the KK 
excitation number. }
\label{diagram}
\end{figure}

Specifically the dominant part of $\Omg_{lX}$ when 
$e^{-2c_l\Re T^{I'}}\gg 1$ and $e^{-2c_X\Re T^{I'}}\ll 1$ 
comes from diagrams depicted in Fig.~\ref{diagram}. 
The internal line in this figure corresponds to 
the $n$-th KK mode of $V^{I'}$, which should be integrated out. 
The effective gauge couplings~$g_l^{(n)}$ 
and $g_X^{(n)}$ are defined as 
\bea
 g_l^{(n)} \eql c_l\int_0^{\pi R}\dr y\;\brc{f_l(y)}^2f_V^{(n)}(y), 
 \nonumber\\
 g_X^{(n)} \eql c_X\int_0^{\pi R}\dr y\;\brc{f_X(y)}^2f_V^{(n)}(y), 
 \label{eff_g}
\eea
where $f_l(y)$, $f_X(y)$ and $f_V^{(n)}(y)$ are wave functions 
in the extra dimension for $\cS_l$, $X$ and $V^{I'(n)}$, respectively. 
Thus the contribution in Fig.~\ref{diagram} disappears 
when either of $c_l$ or $c_X$ vanishes. 
In fact, we can easily check from (\ref{cf_Omg}) 
that $\Omg_{lX}$ is reduced to 
\be
 \Omg_{lX} = \hat{\cN}^{1/3}(\Re T)\frac{1-e^{-2c_X\Re T^1}}
 {3c_X\Re T^1}, 
\ee
when $c_l=0$. 
This is the same form as the single modulus case~(\ref{Omg_single}) 
with $c_l=0$. 
Namely the additional contribution corresponding to Fig.~\ref{diagram} 
disappears. 

Now let us consider the flavor dependence of the contribution 
from Fig.~\ref{diagram}, 
which appears through the $c_l$-dependence of $g_l^{(n)}$. 
The wave function~$f_l(y)$ has an exponential profile 
whose power is proportional to $c_l$. 
We focus on the situation in which $e^{-2c_l\Re T^{I'}}\gg 1$. 
Then $f_l(y)$ is localized around $y=\pi R$. 
In contrast, $f_V^{(n)}$ vanishes at the boundaries 
because $V^{I'}$ are odd under the orbifold parity, 
which means that $f_V^{(n)}$ behaves as a linear function 
around $y=\pi R$. 
Thus the $c_l$-dependence of $g_l^{(n)}$ in (\ref{eff_g}) 
is estimated as $\cO(1/c_l)$ when $f_l^{(n)}$ is localized 
around $y=\pi R$ strongly enough. 
Therefore the flavor dependence of 
the effective coupling~$g_l^{(n)}$ is cancelled and 
the contribution from Fig.~\ref{diagram} becomes flavor universal. 

Finally we comment on the physical degrees of freedom of 
the vector multiplets that contribute to $\Omg_{lX}$. 
Suppose $n_V'$ vector multiplets~$\bdm{V}^{I'}=(V^{I'},\Sgm^{I'})$ 
($I'=1,\cdots,n_V'$). 
Then $n_V'$ moduli~$T^{I'}$ come out from $\Sgm^{I'}$, and 
the remaining degrees of freedom in $\Sgm^{I'}$ are absorbed 
into $V^{I'}$ as longitudinal components of the massive KK vector multiplets. 
However, not all $V^{I'}$ are independent degrees of freedom. 
For example, only $n_V'-1$ gauginos are independent because of 
the gauge-fixing condition for the $\bdm{S}$-supersymmetry in 
the superconformal algebra, that is,\footnote{
To simplify the discussion, here we consider a case that there are 
no $\bdm{V}^{I''}$, namely, $n_V'=n_V$. } 
\be
 \hat{\cN}_{I'}(M)\lmd^{I'} = 0, 
\ee
where $M^{I'}$ and $\lmd^{I'}$ are the gauge scalars and the gauginos. 
As for the vector components of $V^{I'}$, 
one of their combination is identified with the graviphoton, 
which belongs to the 5D SUGRA multiplet. 
It has been noticed that the 5D SUGRA multiplet does not generate 
the contact interactions between $\cS_l$ and $X$~\cite{Luty:2001}. 
In fact, only $n_V'-1$ equations are independent among 
the equations of motion for $V^{I'}$ as we can see from (3.24) 
of Ref.\cite{Abe:2006eg}. 
As a result, the number of the independent $V^{I'}$ that 
contribute to $\Omg_{lX}$ is $n_V'-1$. 
Therefore the contribution from Fig.~\ref{diagram} exists only 
in the multi moduli case.

\section{Moduli stabilization and flavor structure} \label{moduli_stb}
In this section, we investigate the stabilization of the moduli 
and the flavor structure of the soft SUSY breaking terms 
in a specific model. 

\subsection{A model for the hidden sector}
For the moduli stabilization and SUSY breaking, 
we introduce the following boundary superpotentials 
in addition to the Yukawa couplings in (\ref{bd_Yukawa}).  
\bea
 W^{(0)} \eql J_0H+W^{(0)}_{\rm SB}(\Psi), \nonumber\\
 W^{(\pi)} \eql -J_\pi H, 
\eea
where $J_0$ and $J_\pi$ are constants, 
$W^{(0)}_{\rm SB}$ denotes terms relevant for SUSY breaking, 
and $H,\Psi^{a_{\rm SB}}\in\{S_\alp\}$, where 
the index~$a_{\rm SB}$ runs over hypermultiplets in the SUSY breaking sector. 
Then, including the nonperturbative effects such as gaugino condensations, 
the 4D effective superpotential is obtained as 
\bea
 W \eql \brkt{J_0-J_\pi e^{-c_H T^2}}H
 +W^{(0)}_{\rm SB}(\Psi)
 +W^{\rm (np)},  \label{4Deff_sp}
\eea
where $W^{\rm (np)}$ denotes terms coming from the nonperturbative effects 
and is assumed to have a form of 
\be
 W^{\rm (np)} = c-Ae^{-aT^1},  \label{W_np}
\ee
where $A=\cO(1)$, $a=\cO(4\pi^2)$ and $\ln c^{-1}=\cO(4\pi^2)$. 
The $T^1$-dependent (constant) term originates from, e.g., 
a bulk zero-mode (boundary) gaugino condensation. 
The effect of the tadpole terms 
in $W^{(\vtht)}$ ($\vtht=0,\pi$) is discussed in Ref.~\cite{Maru:2003mq} 
in the single modulus case, and they stabilize the (radius) modulus 
at a supersymmetric Minkowski vacuum. 
We will see that the first term of (\ref{4Deff_sp}) stabilizes 
$T^2$ just like in the single modulus case. 
We can choose the O'Raifeartaigh model~\cite{ORft} 
as the SUSY breaking sector~$W^{(0)}_{\rm SB}$, for example. 
It is reduced to 
the Polonyi-type superpotential after heavy modes are integrated out. 
Namely, below the mass scale of the heavy modes~$\Lmd$, 
\be
 W^{(0)}_{\rm SB}(\Psi) \to \mu_X^2 X, 
\ee
where $X$ is one of $\Psi^{a_{\rm SB}}$ that remains at low energies. 
The constant~$\mu_X$ is supposed to be around the TeV scale. 
When the heavy modes are integrated out, the K\"{a}hler potential 
also receives the following correction at one-loop \cite{OKKLT}. 
\be
 \Dlt K = -\frac{Z^{(1)}}{\Lmd^2}\abs{X}^4, \label{Dlt_K}
\ee
where $Z^{(1)}=\cO(1)$ is a constant.\footnote{
When the SUSY breaking sector is introduced both in $W^{(0)}$ 
and $W^{(\pi)}$, $Z^{(1)}$ depends on the moduli~$T^{I'}$. } 

Therefore the effective superpotential below $\Lmd$ becomes 
\bea
 W \eql \brkt{J_0-J_\pi e^{-c_HT^2}}H+c-Ae^{-aT^1}+\mu_X^2X+\cdots,  
\eea
where the ellipsis denotes irrelevant terms to the moduli stabilization 
and SUSY breaking, such as the Yukawa couplings for the MSSM fields. 
From (\ref{eff_def_fcn}) and (\ref{Dlt_K}), the effective K\"{a}hler 
potential~$K$ is 
\bea
 K \eql -3\ln\brkt{-\frac{\Omg}{3}}+\Dlt K \nonumber\\
 \eql -\ln\hat{\cN}+Z_H(\Re T^2)\abs{H}^2+Z_X(\Re T^1)\abs{X}^2 \nonumber\\
 &&+\sum_l Z_l(\Re T)\abs{\cS_l}^2-\frac{Z^{(1)}}{\Lmd^2}\abs{X}^4
 +\cdots,
\eea
where
\bea
 Z_H \defa \frac{1-e^{-2c_H\Re T^2}}{c_H\Re T^2}, \;\;\;
 Z_X \equiv \frac{1-e^{-2c_X\Re T^1}}{c_X\Re T^1}. 
 \label{def:Zs}
\eea
As we will see in the next subsection, the one-loop correction~$\Dlt K$ 
is necessary to obtain a small VEV of $X$.

\subsection{Moduli stabilization and SUSY breaking}
Now we search for a vacuum of the model by solving the minimization 
condition for the scalar potential, which is obtained by the formula, 
\be
 V=e^K\brc{K^{A\bar{B}}D_AWD_{\bar{B}}\bar{W}-3\abs{W}^2}, 
 \label{def:V}
\ee
where $D_AW\equiv W_A+K_A W$. 
The indices~$A,B$ run over 
all the chiral multiplets in the effective theory, 
but it is enough to take $A,B=T^1,T^2,H,X$ in the following 
calculations since the MSSM multiplets do not contribute to 
the moduli stabilization nor SUSY breaking. 
The lower index~$A$ of each function denotes 
derivatives of it for $A$. 

Following Ref.~\cite{AHKO,single_modulus},  
let us define a ``reference point'' which seems to be close 
to the genuine stationary point of the scalar potential. 
We define the reference point~$(T^1,T^2,H,X)|_0$ such that 
the following conditions are satisfied there: 
\bea
 D_{T^1}W \eql aAe^{-aT^1}+K_{T^1}W = 0, \nonumber\\
 D_{T^2}W \eql c_Hw_\pi e^{-c_HT^2}H+K_{T^2}W = 0, \nonumber\\
 D_HW \eql J_0-J_\pi e^{-c_HT^2}+K_HW = 0, \nonumber\\
 V_X \eql 0.  \label{def:ref_pt}
\eea
Here we assume that 
\bea
 &&1 > M_H \equiv \left.c_HJ_\pi e^{-c_HT^2}\right|_0 
 \gg \left.aAe^{-aT^1}\right|_0 = \cO(\mu_X^2),  \nonumber\\
 &&\Re T^1|_0,\,\Re T^2|_0 = \cO(1), \;\;\;
 \abs{X}|_0 \ll 1.  \label{assumptions}
\eea
From the first two conditions in (\ref{def:ref_pt}), we obtain 
\bea
 H|_0 \eql \left.\frac{K_{T^2}}{K_{T^1}}
 \frac{aAe^{-aT^1}}{c_HJ_\pi e^{-c_HT^2}}\right|_0
 = \cO\brkt{\frac{\mu_X^2}{M_H}}, \nonumber\\
 W|_0 \eql \left.-K_{T^1}^{-1}aAe^{-aT^1}\right|_0 = \cO(\mu_X^2). 
 \label{W:ref_pt}
\eea
From the third condition, 
\be
 \left.J_0-J_\pi e^{-c_HT^2}\right|_0 \simeq \left.-Z_H\bar{H}W\right|_0 
 = \cO\brkt{\frac{\mu_X^4}{M_H}} \ll \cO(\mu_X^2). 
\ee
We have assumed that $c_H=\cO(1)>0$, \ie, $Z_H\simlt\cO(1)$. 
Thus, the values of the moduli at the reference point are determined as 
\bea
 T^1|_0 \sima \left.\frac{1}{a}\ln\frac{A(1-K_{T^1}^{-1}a)}{c+\mu_X^2X}
 \right|_0, \nonumber\\
 T^2|_0 \sima \frac{1}{c_H}\ln\frac{J_\pi}{J_0}. 
 \label{T:ref_pt}
\eea
The first equation is consistent with 
(\ref{assumptions}) since $\ln(A/c)=\cO(4\pi^2)$. 
The second equation is similar to the relation in Ref.~\cite{Maru:2003mq} 
if $T^2|_0$ is replaced by the VEV of the radion. 
Besides, $T^2$ and $H$ have 
a large supersymmetric mass of order $\cO(M_H)$, just like 
the situation in Ref.~\cite{Maru:2003mq}. 

From (\ref{def:V}) and (\ref{def:ref_pt}), we obtain 
\bea
 V|_0 \eql e^K\brc{K^{X\bar{X}}\abs{D_XW}^2-3\abs{W}^2} \nonumber\\
 \sima e^K\brc{Z_X^{-1}\abs{\mu_X}^4-3\abs{W}^2}. 
\eea
Thus we can make $V|_0=0$ by tuning $\mu_X$ as 
\be
 \abs{\mu_X}^4 \simeq 3Z_X\abs{W}^2. 
 \label{cond:zero_cc}
\ee
This is consistent with the second equation in (\ref{W:ref_pt}) 
when $c_X>0$. 

The value of $X|_0$ is determined by the last condition 
in (\ref{def:ref_pt}). 
Under the condition that $V|_0=0$, 
$V_X$ at the reference point is estimated as 
\bea
 e^{-K}V_X \eql \der_XK^{X\bar{X}}\abs{D_XW}^2
 +K^{A\bar{X}}\der_XD_AWD_{\bar{X}}\bar{W}
 +K^{X\bar{B}}D_XW\der_XD_{\bar{B}}\bar{W}-3\bar{W}W_X \nonumber\\
 \sima \der_XK^{X\bar{X}}\abs{D_XW}^2
 +K^{X\bar{X}}D_XW\der_XD_{\bar{X}}\bar{W}-3\bar{W}W_X \nonumber\\
 \sima \frac{4\abs{\mu_X}^4Z^{(1)}}{\Lmd^2Z_X^2}\bar{X}
 -2\mu_X^2\bar{W}. 
\eea
We have omitted the symbol~$|_0$. 
Therefore, we obtain 
\be
 X|_0 \simeq \left.\frac{\Lmd^2Z_X^2W}{2\mu_X^2Z^{(1)}}\right|_0 
 = \cO(\Lmd^2). 
 \label{X:ref_pt}
\ee
The last assumption in (\ref{assumptions}) can be satisfied 
by assuming that $\Lmd\ll M_{\rm Pl}=1$. 

The true vacuum is represented by 
\be
 \vev{A} = A|_0+\dlt A, 
\ee
where $A=T^1,T^2,H,X$. 
Since $T^2$ and $H$ have a large mass, 
$\dlt T^2$ and $\dlt H$ are negligible~\cite{Abe:2006xi} 
if we take $M_H$ as around $M_{\rm GUT}$ or so.\footnote{
They are estimated as $\dlt T^2,\dlt H=\cO(\mu_X/M_H)$. }
Thus $T^2$ and $H$ can be replaced with their VEVs, 
which are equal to the values at the reference point 
in the following calculation. 
Now we find a true vacuum by solving the minimization conditions:
\be
 V_{T^1}=V_X=0. \label{minimize_V}
\ee 
We can evaluate the derivatives of the potential as 
\bea
 e^{-K}V_{T^1} \sima \left.K^{T^1\bar{T}^1}\abs{W_{T^1T^1}}^2\right|_0
 \dlt\bar{T}^1+\left.\bar{\mu}_X^2K^{T^1\bar{T}^{I'}}K_{\bar{T}^{I'}}
 W_{T^1T^1}\right|_0\dlt\bar{X}+\left.3K_{T^1}\abs{W}^2\right|_0, \nonumber\\
 e^{-K}V_X \sima \left.\mu_X^2K_{T^1}\bar{W}\right|_0\dlt T^1
 +\left.\mu_X^2K^{T^{I'}\bar{T}^1}K_{T^{I'}}\overline{W_{T^1T^1}}
 \right|_0\dlt\bar{T}^1
 +\left.\frac{2\mu_X^2\bar{W}}{\bar{X}}\right|_0\dlt\bar{X}. 
\eea
Here we have used (\ref{cond:zero_cc}) and 
\be
 \abs{X}|_0 \ll 1, \quad
 \abs{W_{T^1T^1}}^2 = \cO(a\mu_X^2) \gg \abs{W},\abs{W_{T^1}} 
 = \cO(\mu_X^2).   
\ee
Using the relation followed from the first condition in (\ref{def:ref_pt}): 
\be
 W_{T^1T^1} = -a^2Ae^{-aT^1} = aK_{T^1}W, 
\ee
we obtain by solving (\ref{minimize_V}),
\bea
 \dlt T^1 \sima \left.-\frac{3}{a^2K^{T^1\bar{T}^1}K_{T^1}}\right|_0, 
 \nonumber\\
 \dlt X \sima \left.\frac{3X\brkt{K^{T^1\bar{T}^1}K_{\bar{T}^1}
 +K^{T^1\bar{T}^2}K_{\bar{T}^2}}}
 {2aK^{T^1\bar{T}^1}}\right|_0. 
\eea
Thus we can evaluate $D_AW$ ($A=T^1,T^2,H,X$) as 
\bea
 D_{T^1}W \sima \left.-\frac{3W}{aK^{T^1\bar{T}^1}}\right|_0 
 = \cO\brkt{\frac{\mu_X^2}{a}}, \nonumber\\
 D_{T^2}W \sima \left.-\frac{3(2K_{T^1T^2}-K_{T^1}K_{T^2})W}
 {a^2K^{T^1\bar{T}^1}K_{T^1}}\right|_0 
 = \cO\brkt{\frac{\mu_X^2}{a^2}}, \nonumber\\
 D_HW \sima 0, \nonumber\\
 D_XW \sima \mu_X^2. 
\eea
The $F$-terms are calculated from these 
by the formula~$F^A=-e^{K/2}K^{A\bar{B}}\overline{D_BW}$. 

In the case that the norm function is chosen as (\ref{norm1}), 
the inverse of the K\"{a}hler metric is 
\be
 K^{A\bar{B}} \simeq \begin{pmatrix} 
 \frac{2\brc{2(\Re T^1)^3-(\Re T^2)^3}}{3\Re T^1} & 2(\Re T^1)(\Re T^2) 
 & & \\
 2(\Re T^1)(\Re T^2) & \frac{2\brc{-(\Re T^1)^3+2(\Re T^2)^3}}{3\Re T^2} 
 & & \\ & & Z_H^{-1} & \\ & & & Z_X^{-1} 
 \end{pmatrix}. 
\ee
Thus the $F$-terms are estimated as 
\bea
 F^{T^1} \sima \frac{3\bar{W}}{a\brc{(\Re T^1)^3+(\Re T^2)^3}^{1/2}}
 = \cO\brkt{\frac{\mu_X^2}{a}}, \nonumber\\
 F^{T^2}
 \sima \frac{9(\Re T^1)^2(\Re T^2)\bar{W}}
 {a\brc{(\Re T^1)^3+(\Re T^2)^3}^{1/2}
 \brc{2(\Re T^1)^3-(\Re T^2)^3}}
 = \cO\brkt{\frac{\mu_X^2}{a}},  \nonumber\\
 F^H \sima 0, \nonumber\\
 F^X 
 \sima -\frac{c_X\bar{\mu}_X^2\Re T^1}{\brc{(\Re T^1)^3+(\Re T^2)^3}^{1/2}}
 = \cO(\mu_X^2), 
 \label{Fs:eg1}
\eea
where we used the relation (\ref{cond:zero_cc}). 
Therefore, the relation~(\ref{F:assumption}) holds in this model. 

In the case that the norm function is chosen as (\ref{norm2}), 
the inverse of the K\"{a}hler metric becomes diagonal, \ie, 
\be
 K^{A\bar{B}} \simeq \diag\brkt{2(\Re T^1)^2,4(\Re T^2)^2,Z_H^{-1},Z_X^{-1}}, 
\ee
and the $F$-terms are estimated as 
\bea
 F^{T^1}  \sima 
 \frac{3\bar{W}}{a(\Re T^1)(\Re T^2)^{1/2}} 
 = \cO\brkt{\frac{\mu_X^2}{a}},  \nonumber\\
 F^{T^2} \sima \frac{3(\Re T^2)^{1/2}\bar{W}}{a^2(\Re T^1)^3} 
 = \cO\brkt{\frac{\mu_X^2}{a^2}},  \nonumber\\
 F^H \sima 0, \nonumber\\
 F^X 
 \sima -\frac{c_X\bar{\mu}_X^2}{(\Re T^2)^{1/2}} = \cO(\mu_X^2), 
 \label{Fs:eg2}
\eea
where we again used the relation (\ref{cond:zero_cc}). 
Thus (\ref{F:assumption}) holds. 
In this case, further hierarchy exist between $F^{T^1}$ and $F^{T^2}$. 

Finally we comment that the above moduli stabilization 
with hierarchical $F$-terms is motivated by a type IIB flux 
compactification~\cite{Choi:2004sx}, 
where a size modulus is stabilized by a nonperturbative 
effect, while complex structure moduli are stabilized 
at a high scale by a flux induced superpotential. 
In our model, the terms with $J_0$ and $J_\pi$ in (\ref{4Deff_sp}) 
play a similar role to the flux induced superpotential 
if $T^1$ and $(T^2,H)$ are identified with the size modulus and 
the shape moduli respectively. 

\subsection{A model for the visible sector and sparticle spectrum} \label{num_calc}
Now we study some phenomenological consequences such as the 
hierarchical Yukawa matrices and the soft SUSY breaking parameters. 
The visible (MSSM) sector consists of 
\begin{eqnarray}
({\cal Q}_i, {\cal U}_i, {\cal D}_i) &:& \textrm{quark supermultiplets} 
 \ (i=1,2,3), \nonumber\\
({\cal L}_i, {\cal E}_i) &:& \textrm{lepton supermultiplets} \ (i=1,2,3), 
 \nonumber\\
({\cal H}_u, {\cal H}_d) &:& \textrm{Higgs supermultiplets}. 
\end{eqnarray}
\ignore{
Before that we summarize the moduli stabilization. 
The moduli $T^1$ and $T^2$ are stabilized at (\ref{T:ref_pt}) 
by a nonperturbative effect (\ref{W_np}) 
and a superpotential (SUSY) mass term with $H$, respectively. 
On the other hand, the SUSY breaking field $X$ is stabilized 
at (\ref{X:ref_pt}) by a combination of a one-loop SUSY 
breaking mass and a gravitational effect. 
}
As we saw in the previous subsection, 
the $F$-terms in the SUSY breaking sector have a hierarchical structure.  
From (\ref{Fs:eg1}) or (\ref{Fs:eg2}) with (\ref{cond:zero_cc}), 
we obtain 
\begin{eqnarray}
\frac{F^{T^1}}{\Re T^1} 
 \sima \frac{3m_{3/2}}{a\Re T^1} 
 = \cO\brkt{\frac{m_{3/2}}{4\pi^2}}, \qquad 
F^X \simeq -\sqrt{\frac{3}{Z_X}}m_{3/2} = \cO(m_{3/2}), 
\end{eqnarray}
and 
\be
 \frac{F^{T^2}}{\Re T^2} \sim \begin{cases}
 {\displaystyle \frac{F^{T^1}}{\Re T^1}} & \mbox{for $\cN(X)=(X^1)^3+(X^2)^3$} 
 \\*[5pt]
 {\displaystyle \frac{1}{4\pi^2}\frac{F^{T^1}}{\Re T^1}} & \mbox{for $\cN(X)=(X^1)^2X^2$} 
 \end{cases}, 
 \label{F_T2}
\ee
where $m_{3/2}\equiv e^{K/2}W|_0$ is the gravitino mass. 
We have assumed that $W|_0$ and $\mu_X$ are real and positive, 
for simplicity. 
These relations hold also in the case that 
VEVs of the moduli take values of $\cO(10)$ 
as long as $a\Re T^{I'}=\cO(4\pi^2)$ ($I'=1,2$). 

For the numerical estimations in the following, 
we assume that $a=T^1|_0=T^2|_0=2\pi$. 
Besides we focus on the second case in (\ref{F_T2}). 
Then contributions from $F^{T^2}$ can be neglected 
due to the suppression factor~$1/(4\pi^2)$. 
We take $M_{SB}\equiv F^{T^1}/(2\Re T^1)$ as a reference scale 
of SUSY breaking. 
The gravitino mass is then expressed as 
$m_{3/2}\simeq (8\pi^2/3)M_{SB}$.

\ignore{
\begin{eqnarray}
\frac{F^{T^2}}{T^2+\bar{T}^2} 
\approx \frac{F^{T^1}}{T^1+\bar{T}^1} 
\quad \textrm{for} \quad 
{\cal N} = (M^1)^3 + (M^2)^3, 
\end{eqnarray}
or 
\begin{eqnarray}
\frac{F^{T^2}}{T^2+\bar{T}^2} 
\approx \frac{1}{4\pi^2} \frac{F^{T^1}}{T^1+\bar{T}^1} 
\quad \textrm{for} \quad 
{\cal N} = (M^1)^n(M^2)^{3-n} \quad (n=1,2). 
\end{eqnarray}
Then, in the following analysis of the visible sector, 
we parameterize the SUSY breaking order parameters 
by ($\alpha$, $\beta$, $\gamma$) as 
\begin{eqnarray}
\frac{F^{T^1}}{T^1+\bar{T}^1} \equiv M_{SB}, \quad 
\frac{F^{T^2}}{T^2+\bar{T}^2} = \beta M_{SB}, \quad 
m_{3/2} = 4\pi^2 \alpha M_{SB}, 
\label{fratios}
\end{eqnarray}
and 
\begin{eqnarray}
F^X = -\gamma \sqrt{3Z_X} c_X \Re T^1|_0\,m_{3/2}, \quad 
\frac{F^\phi}{\phi} = m_{3/2}-\frac{1}{3}K_AF^A 
\quad (A=T^1,T^2,X), \quad 
\end{eqnarray}
where $M_{SB}=F^{T^1}/(T^1+\bar{T}^1)$ is adopted as a 
reference scale of the SUSY breaking. 
As shown in the previous section, 
the natural orders of the parameters are 
\begin{eqnarray}
(\alpha, \beta, \gamma) 
&\simeq& (1,1,1) 
\quad \textrm{or} \quad 
(1,1/4\pi^2,1) 
\end{eqnarray}
depending on the norm function ${\cal N}$. 
For the latter case, $F^{T^2}/(T^2+\bar{T}^2)$ is suppressed 
by $1/(4 \pi^2)^2$ compared with the gravitino mass $m_{3/2}$, 
and then the contributions from $F^{T^2}$ is negligible, i.e., 
we can take $\beta=0$ effectively. 
For the numerical estimations in the following, 
the stabilized values of $T^1$ and $T^2$ are assumed as 
$T^1|_0 = T^2|_0 = a = 2 \pi$. 
In Fig.~\ref{mQimSa}, we show the scalar masses of $Q_i$ and $S_\alpha$ 
as a function of $\phi_{Q_i}$ and $\phi_X$ which are defined by 
\begin{eqnarray}
\phi_{Q_i} 
= \frac{\Re T^1|_0}{\ln \varepsilon^{-1}}\,c_{Q_i}, \qquad 
\phi_{S_\alpha} 
= \frac{\Re T^1|_0}{\ln \varepsilon^{-1}}\,c_{S_\alpha}, \qquad 
\phi_X = \frac{\Re T^1|_0}{\ln \varepsilon^{-1}}\,c_X. 
\end{eqnarray}
where $\varepsilon \simeq 0.22$ is the Cabibbo angle. 
From the figures, we find that, for $\phi_X>0$, 
the scalar mass square is positive and almost charge 
(i.e., flavor) independent for $S_\alpha$ 
with negative $\phi_{S_\alpha}$. 
\begin{figure}[t]
\begin{center}
\begin{minipage}{0.4\linewidth}
\begin{center}
\epsfig{file=mq1.eps,width=\linewidth} 
\end{center}
\end{minipage}
\hspace{20pt}
\begin{minipage}{0.4\linewidth}
\begin{center}
\epsfig{file=mq2.eps,width=\linewidth} 
\end{center}
\end{minipage}
\end{center}
\caption{Charge dependences of the scalar masses. 
The perpendicular axes represents the magnitude of 
the square root of the scalar mass square multiplied by its sign.
The parameters are chosen as $(\alpha,\beta,\gamma)=(1,0,1)$.} 
\label{mQimSa}
\end{figure}
}

We assume an approximate global $U(1)_R$-symmetry that is 
responsible for the dynamical SUSY breaking. 
We assign $R(X)=2$, $R({\cal H}_u)=R({\cal H}_d)=1$, 
$R({\cal Q}_i)=R({\cal U}_i)=R({\cal D}_i)
=R({\cal L}_i)=R({\cal E}_i)=1/2$ 
where $R(\Phi)$ is the R-charge of $\Phi$, 
and assume that the R-symmetry is broken only by 
the nonperturbative effects $W^{(\textrm{np})}$. 
In this case, the holomorphic Yukawa couplings and the $\mu$-term 
in the 4D effective superpotential as well as the gauge kinetic 
functions are independent of $X$. 
We further assume that the Yukawa couplings and the 
$\mu$-term originate from only the $y=0$ boundary. 
\ignore{
In the following, we assume such a case, that is, all the holomorphic 
Yukawa couplings and the $\mu$-term are constant, and the visible 
gauge kinetic functions are (linear) functions of $T^1$ and $T^2$ 
but not $X$. 
}
Then they are parameterized as  
\begin{eqnarray}
f_r &=& k_{r1}T^1+k_{r2}T^2, \nonumber\\
W_{\rm MSSM} &=& \mu {\cal H}_u {\cal H}_d
+\lambda^u_{ij} {\cal H}_u {\cal Q}_i {\cal U}_j
+\lambda^d_{ij} {\cal H}_d {\cal Q}_i {\cal D}_j
+\lambda^e_{ij} {\cal H}_d {\cal L}_i {\cal E}_j, 
\label{mssmsp}
\end{eqnarray} 
where $r=1,2,3$ for $U(1)_Y$, $SU(2)_L$, $SU(3)_C$ respectively, 
and $k_{r1}$, $k_{r2}$, $\mu$, 
$\lambda^{u,d,e}_{ij}$ are constants.  
These constants are understood as values at the compactification scale, 
which is close to $M_{\rm Pl}$. 
The constant~$k_{r2}$ is related to $k_{r1}$ by the condition that 
the three gauge couplings are unified to a definite value 
at $M_{\rm GUT}$. 
In the following, we neglect the RGE running between $M_{\rm Pl}$ 
and $M_{\rm GUT}$. 

The hierarchical structure of the physical Yukawa couplings 
$y_{lmn}$ defined in (\ref{fml_softSB}) 
are generated with certain choices of $V^1$- or $V^2$-charges, 
$c_l$, for the visible matter multiplets 
$\cS_l=({\cal Q}_i,{\cal U}_i,{\cal D}_i,{\cal L}_i,{\cal E}_i,
{\cal H}_u,{\cal H}_d)$, which appear nontrivially in the 
superspace wavefunctions $Y_l$ shown in (\ref{Omg_expand}). 
In this case, as discussed previously, 
the tachyonic sfermion masses would be avoided by suitably 
gauging $\cS_l$ by either the $Z_2$-odd $U(1)$ vector multiplet 
$V^1$ or $V^2$. 
In order to obtain a realistic pattern of Yukawa matrices 
without inducing tachyonic sfermion masses, 
we adopt the gauging for quarks and leptons as 
(see Example~2 in Sec.~\ref{flavor_soft}) 
\begin{eqnarray}
{\cal Q}_{i=3}, {\cal U}_{i=3} \in \{Q_i\} 
\ \textrm{gauged by} \ V^1, \label{3rdgngauging} \\
{\cal Q}_{i \ne 3}, {\cal U}_{i \ne 3}, {\cal D}_i, 
{\cal L}_i, {\cal E}_i \in \{S_\alpha\} 
\ \textrm{gauged by} \ V^2, 
\end{eqnarray}
and employ a charge assignment given by 
Refs.~\cite{CKKK,ACJO}, that is, 
\begin{eqnarray}
 c_{\cQ_i}\Re T \eql (-4.5,-3,3), \quad 
 c_{{\cal U}_i}\Re T=(-7.5,-3,3), \nonumber\\ 
 c_{\cD_i}\Re T^2 \eql (-4.5,-4.5,-3), \quad
 c_{\cL_i}\Re T^2=(-4.5,-4.5,-1.5), \nonumber\\
 c_{{\cal E}_i}\Re T^2 \eql (-6,-1.5,-1.5), \label{quarkcharges}
\end{eqnarray}
where $\Re T$ in the first line represents $\Re T^2$ 
for the first two generations and $\Re T^1$ for the third generation. 
For the Higgs multiplets, we take the gauging 
\begin{eqnarray}
{\cal H}_u  \in \{S_\alpha\}, \quad
{\cal H}_d  \in \{Q_i\}, 
\end{eqnarray}
and a charge assignment, 
\begin{eqnarray}
 c_{{\cal H}_u}\Re T^2=3, \quad 
 c_{{\cal H}_d}\Re T^1=12. 
\end{eqnarray}
Note that only $\cH_u$ is tachyonic with this gauging, 
which is sufficient condition for the electroweak 
symmetry breaking even when the gaugino masses are 
tiny compared with the scalar masses, for which the 
radiative electroweak breaking might be impossible. 
We choose $c_{{\cal H}_d}$ as a larger value 
than others to allow for a mildly large value of 
\begin{eqnarray}
\tan \beta \equiv 
\langle {\cal H}_u \rangle / \langle {\cal H}_d \rangle = 5, 
\label{tanbeta}
\end{eqnarray}
which is adopted in the following numerical evaluations. 

\ignore{
\begin{figure}[t]
\begin{center}
\begin{minipage}{0.5\linewidth}
\begin{minipage}{\linewidth}
\begin{center}
\epsfig{file=gmatmg.eps,width=\linewidth} 
\end{center}
\end{minipage}
\begin{minipage}{\linewidth}
\begin{center}
\epsfig{file=smatmg.eps,width=\linewidth} 
\end{center}
\end{minipage}
\end{minipage}
\begin{minipage}{0.15\linewidth}
\begin{center}
\epsfig{file=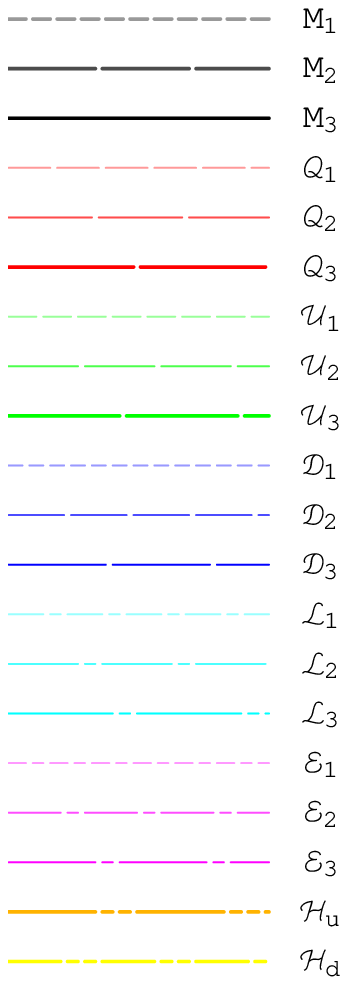,width=\linewidth}
\end{center}
\end{minipage}
\end{center}
\caption{The gaugino masses $M_r$ at $M_{GUT}$ 
as a function of $k \equiv k_1=k_2=k_3$ and 
the scalar masses $m^2_{{\cal Q}_l}$ at $M_{GUT}$ 
as a function of $\phi_X$, normalized by $M_{SB}$. 
The parameters are chosen as $(\alpha,\beta,\gamma)=(1,0,1)$.}
\label{gspectrum}
\end{figure}
}

\ignore{
With the above data for the parameters, we can 
explicitly calculate the gaugino masses, squark/slepton/Higgs 
masses and A-terms as well as the physical Yukawa couplings 
at the cutoff scale, which is taken to be $M_{GUT}$. 
For the parameters $(\alpha,\beta,\gamma)=(1,0,1)$, 
The gaugino masses $M_r$ at $M_{GUT}$ 
as a function of $k \equiv k_1=k_2=k_3$ 
and the scalar masses $m^2_{\cS_l}$ at $M_{GUT}$ 
as a function of 
$\phi_X=\frac{{\rm Re}\,T^1|_0}{
\ln (\varepsilon^{-1})}\,c_X$ 
normalized by $M_{SB}$ 
are shown in Figs.~\ref{gspectrum} and \ref{gspectrum2}. 
The A-terms are evaluated at $M_{GUT}$ as 
\begin{eqnarray}
\frac{A_u}{M_{SB}} \simeq 
\left( \begin{array}{ccc}
-1 & 
-1 & 
-\varepsilon \\
-1 & 
-1 & 
-\varepsilon \\
-\varepsilon & 
-\varepsilon & 
1 
\end{array} \right), \quad 
\frac{A_d}{M_{SB}} \simeq 
\left( \begin{array}{ccc}
-\varepsilon & 
-\varepsilon & 
-\varepsilon \\
-\varepsilon & 
-\varepsilon & 
-\varepsilon \\
1 & 
1 & 
1 
\end{array} \right), \\
\frac{A_e}{M_{SB}} \simeq 
\left( \begin{array}{ccc}
-1 & 
-1 & 
-1 \\
-1 & 
-1 & 
-1 \\
-1 & 
-1 & 
-1 
\end{array} \right). 
\end{eqnarray}
Note that these numerical values of the soft SUSY 
breaking parameters include anomaly mediated contributions 
as well as the tree-level contributions, which can be 
sizable in gaugino masses and the A-terms. 
}

The physical Yukawa matrices are found as 
\bea
 y_u \sima \begin{pmatrix} \vep^8 & \vep^5 & \vep^3 \\
 \vep^7 & \vep^4 & \vep^2 \\ \vep^5 & \vep^2 & \vep^0
 \end{pmatrix}, \quad
 y_d \simeq \begin{pmatrix} \vep^5 & \vep^5 & \vep^4 \\
 \vep^4 & \vep^4 & \vep^3 \\ \vep^2 & \vep^2 & \vep^1 
 \end{pmatrix}, \quad
 y_e \simeq \begin{pmatrix} \vep^7 & \vep^4 & \vep^4 \\
 \vep^7 & \vep^4 & \vep^4 \\ \vep^5 & \vep^2 & \vep^2 
 \end{pmatrix}, 
\eea
where $\vep=0.22$ is the Cabibbo angle. 
We have omitted an $\cO(1)$ coefficient~$\lmd_{ij}^{u,d,e}$ 
for each element. 
These matrices realize the observed quark and charged lepton 
masses as well as the Cabibbo-Kobayashi-Maskawa (CKM) matrix 
with $\cO(1)$ values of $\lmd_{ij}^{u,d,e}$. 

\ignore{
\left( \begin{array}{ccc}
\lambda^u_{11} \varepsilon^8 & 
\lambda^u_{12} \varepsilon^5 & 
\lambda^u_{13} \varepsilon^3 \\
\lambda^u_{21} \varepsilon^7 & 
\lambda^u_{22} \varepsilon^4 & 
\lambda^u_{23} \varepsilon^2 \\
\lambda^u_{31} \varepsilon^5 & 
\lambda^u_{32} \varepsilon^2 & 
\lambda^u_{33} \varepsilon^0 
\end{array} \right), \quad 
y_d \simeq 
\left( \begin{array}{ccc}
\lambda^d_{11} \varepsilon^5 & 
\lambda^d_{12} \varepsilon^5 & 
\lambda^d_{13} \varepsilon^4 \\
\lambda^d_{21} \varepsilon^4 & 
\lambda^d_{22} \varepsilon^4 & 
\lambda^d_{23} \varepsilon^3 \\
\lambda^d_{31} \varepsilon^2 & 
\lambda^d_{32} \varepsilon^2 & 
\lambda^d_{33} \varepsilon^1 
\end{array} \right), \\
y_e \simeq 
\left( \begin{array}{ccc}
\lambda^e_{11} \varepsilon^7 & 
\lambda^e_{12} \varepsilon^4 & 
\lambda^e_{13} \varepsilon^4 \\
\lambda^e_{21} \varepsilon^7 & 
\lambda^e_{22} \varepsilon^4 & 
\lambda^e_{23} \varepsilon^4 \\
\lambda^e_{31} \varepsilon^5 & 
\lambda^e_{32} \varepsilon^2 & 
\lambda^e_{33} \varepsilon^2 
\end{array} \right), 
\end{eqnarray}
which can realize observed quark and charged lepton masses 
as well as the Cabibbo-Kobayashi-Maskawa (CKM) matrix 
with certain values of holomorphic Yukawa couplings 
$\lambda^{u,d,e}_{ij}$ of ${\cal O}(1)$ in $W_{MSSM}$. 
\begin{figure}[t]
\begin{center}
\begin{minipage}{0.5\linewidth}
\begin{minipage}{\linewidth}
\begin{center}
\epsfig{file=gmatmg2.eps,width=\linewidth} 
\end{center}
\end{minipage}
\begin{minipage}{\linewidth}
\begin{center}
\epsfig{file=smatmg2.eps,width=\linewidth} 
\end{center}
\end{minipage}
\end{minipage}
\begin{minipage}{0.15\linewidth}
\begin{center}
\epsfig{file=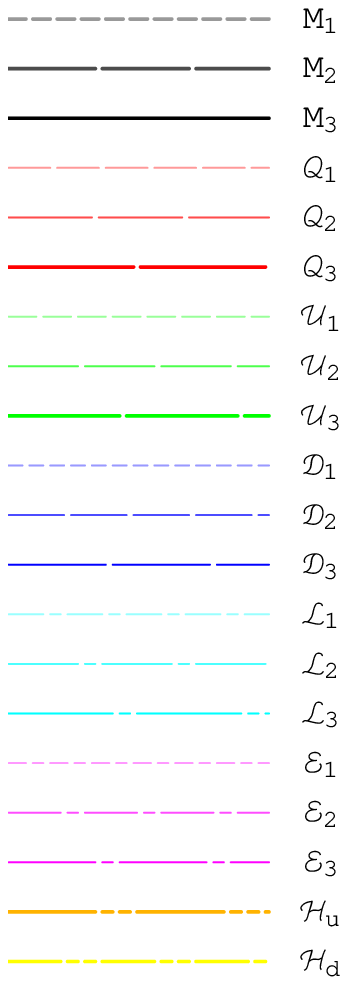,width=\linewidth}
\end{center}
\end{minipage}
\end{center}
\caption{The same plots as Fig.~\ref{gspectrum} 
but with different scales of axes. 
The splitting of the gaugino masses $M_r$ is caused 
by the anomaly mediated contributions. 
The scalar masses $m^2_{{\cal Q}_l}$ with 
$|\phi_{{\cal Q}_l}| \gtrsim 2$ tend to degenerate 
except for 
${\cal Q}_3$, ${\cal U}_3$, ${\cal H}_u$ and ${\cal H}_d$. 
}
\label{gspectrum2}
\end{figure}
}

\begin{figure}[t]
\begin{center}
\begin{minipage}{0.75\linewidth}
\begin{center}
\epsfig{file=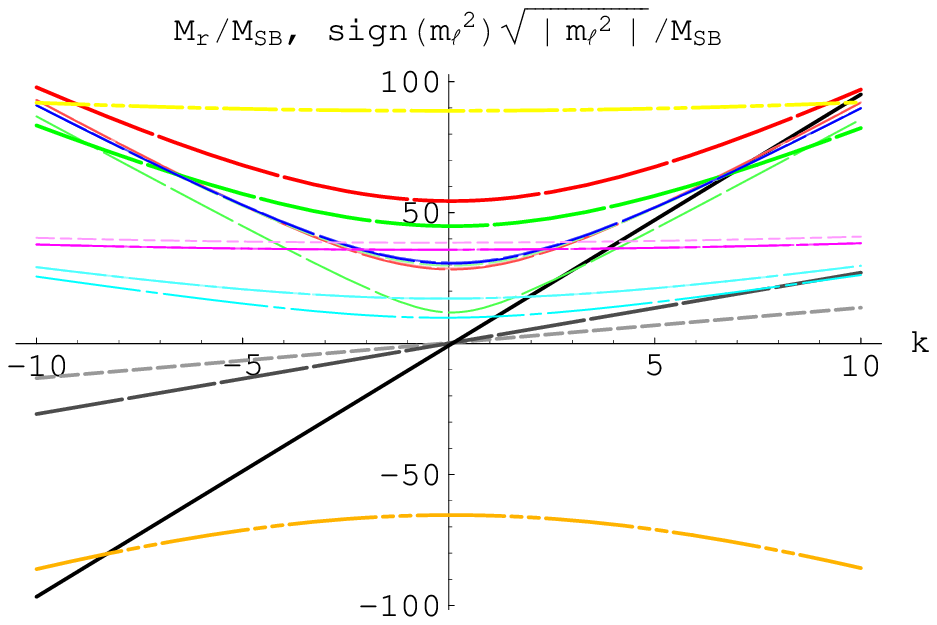,width=\linewidth}
\end{center}
\end{minipage}
\begin{minipage}{0.2\linewidth}
\begin{center}
\epsfig{file=legend_m.eps,width=\linewidth}
\end{center}
\end{minipage}
\end{center}
\caption{The gaugino masses $M_r$ and 
the scalar masses $m^2_l$ at $M_Z$ 
as functions of $k \equiv k_{11}=k_{21}=k_{31}$. 
The parameters are chosen as $c_X\Re T^1=7.5$.}
\label{zspectrum}
\end{figure}

By evaluating one-loop RGEs for MSSM\footnote{
We neglect all Yukawa couplings except for the top Yukawa coupling 
in evaluating MSSM RGEs.} (including the anomaly mediated 
contributions~\cite{Randall:1998uk} which can be sizable 
in gaugino masses and A-terms), we can estimate 
the soft SUSY breaking parameters at $M_Z \simeq 90$ GeV. 
For $c_X\Re T^1=7.5$, the gaugino masses $M_r$ and 
the scalar masses $m^2_{\cS_l}$ at $M_Z$ 
as functions of $k \equiv k_{11}=k_{21}=k_{31}$ 
normalized by $M_{SB}$ 
are shown in Fig.~\ref{zspectrum}. 
From the figure, we find that the gauginos become 
heavier for larger $|k|$, and the gaugino masses 
and the scalar masses are comparable for $|k| \gtrsim 5$. 
For $k=1$, the A-terms are evaluated at $M_Z$ as 
\bea
 \frac{A_u}{M_{SB}} \sima \begin{pmatrix}
 \vep^{-1} & \vep^{-1} & \vep^{-1} \\
 \vep^{-1} & \vep^{-1} & \vep^{-1} \\
 \vep^{-1} & \vep^{-1} & \vep^{-1} \end{pmatrix}, \;
 \frac{A_d}{M_{SB}} \simeq \begin{pmatrix}
 \vep^{-2} & \vep^{-2} & \vep^{-2} \\
 \vep^{-2} & \vep^{-2} & \vep^{-2} \\
 \vep^{-2} & \vep^{-2} & \vep^{-2} \end{pmatrix}, \;
 \frac{A_e}{M_{SB}} \simeq \begin{pmatrix}
 1 & 1 & 1 \\ 1 & 1 & 1 \\ 1 & 1 & 1 \end{pmatrix}. 
\eea
Remark that A-terms in the quark sector are enhanced 
by the radiative corrections mainly from gluinos, 
while there is no enhancement in the lepton sector. 

\ignore{
\begin{eqnarray}
\frac{A_u}{M_{SB}} \simeq 
\left( \begin{array}{ccc}
\varepsilon^{-1} & 
\varepsilon^{-1} & 
\varepsilon^{-1} \\
\varepsilon^{-1} & 
\varepsilon^{-1} & 
\varepsilon^{-1} \\
\varepsilon^{-1} & 
\varepsilon^{-1} & 
\varepsilon^{-1} 
\end{array} \right), \quad 
\frac{A_d}{M_{SB}} \simeq 
\left( \begin{array}{ccc}
\varepsilon^{-2} & 
\varepsilon^{-2} & 
\varepsilon^{-2} \\
\varepsilon^{-2} & 
\varepsilon^{-2} & 
\varepsilon^{-2} \\
\varepsilon^{-2} & 
\varepsilon^{-2} & 
\varepsilon^{-2} 
\end{array} \right), \\ 
\frac{A_e}{M_{SB}} \simeq 
\left( \begin{array}{ccc}
1 & 
1 & 
1 \\
1 & 
1 & 
1 \\
1 & 
1 & 
1 
\end{array} \right). 
\end{eqnarray}
}

Rotating scalar masses and A-terms into the super-CKM basis, 
we can estimate the mass insertion parameters at $M_Z$ 
which are defined by (see, e.g., \cite{Misiak:1997ei,ACJO}) 
\begin{eqnarray}
(\delta^f_{LL})_{ij} \defa
\frac{\left( (V^f_L)^\dagger m_{\tilde{f}_L}^2 V^f_L \right)_{ij}
}{\sqrt{(m_{\tilde{f}_L}^2)_{ii} (m_{\tilde{f}_L}^2)_{jj}}}, \qquad 
(\delta^f_{RR})_{ij} \equiv
\frac{\left( (V^f_R)^\dagger m_{\tilde{f}_R}^2 V^f_R \right)_{ij}
}{\sqrt{(m_{\tilde{f}_R}^2)_{ii} (m_{\tilde{f}_R}^2)_{jj}}}, \nonumber\\
(\delta^f_{LR})_{ij} \defa
\frac{v_f \left( (V^f_L)^\dagger \tilde{A}_f V^f_R \right)_{ij}
-\mu_f \delta_{ij} (m_f)_i 
}{\sqrt{(m_{\tilde{f}_L}^2)_{ii} (m_{\tilde{f}_R}^2)_{jj}}}, \qquad 
(\delta^f_{RL})_{ij} \equiv (\delta^f_{LR})^\dagger_{ij}, 
\end{eqnarray}
where $m_{\tilde{f}_{L,R}}^2$ are the diagonal sfermion mass matrices, 
and $(\tilde{A}_f)_{ij}=(y_f)_{ij} (A_f)_{ij}$ are 
the scalar trilinear couplings. 
The fermion index~$f$ and the sfermion indices~$\tl{f}_{L,R}$ represent 
$f=(u,d,e)$, $\tl{f}_R=(\tl{u},\tl{d},\tl{e})$ 
and $\tl{f}_L=(\tl{q},\tl{q},\tl{l})$, and then 
$v_f=(\sin \beta, \cos \beta, \cos \beta)v$, 
$\mu_f=(\cot \beta, \tan \beta, \tan \beta) \mu$, respectively, 
where $v \simeq 174$ GeV and $\mu$ is determined by the minimization 
condition of the Higgs potential. 
The unitary matrices $V^f_{L,R}$ are defined by 
$(V^f_L)^\dagger y_f V^f_R={\rm diag}\,\left( (m_f)_i/v_f \right)$. 

The mass insertion parameters are severely constrained by 
the experiments of flavor changing processes. 
Among these parameters, $(\delta^e_{LR})_{21}$ 
might have the severest constraint from the observations 
of $\mu \to e \gamma$ processes, and 
$(\delta^d_{LL})_{23}$ 
from $b \to s \gamma$ processes in our model. 
Although $b \to s \gamma$ is less restrictive than 
$\mu \to e \gamma$, the former can be relevant because 
there are large mass splittings between the third generation 
squarks and the first two generations with our 
gauging (\ref{3rdgngauging}) and 
the charge assignment (\ref{quarkcharges}). 
\ignore{
For $M_{SB}=100$ GeV and with certain values of 
$\lambda^{u,d,e}_{ij}$ of ${\cal O}(1)$ in $W_{MSSM}$, 
the parameters $(\delta^e_{LR})_{21}$ and $(\delta^d_{LL})_{23}$ 
are plotted in Fig.~\ref{deltas} 
as a function of $k \equiv k_1=k_2=k_3$. }
For $M_{SB}=100$ GeV and with $\cO(1)$ values of the holomorphic 
Yukawa couplings $\lambda^{u,d,e}_{ij}$, we find 
$|(\delta^e_{LR})_{21}| \lesssim {\cal O}(10^{-6})$ and 
$|(\delta^d_{LL})_{23}(\delta^d_{LR})_{33}| \lesssim {\cal O}(10^{-3})$ 
within $|k| \le 10$. 
Then, roughly speaking, these parameters are typically 
within the allowed region~\cite{ACJO} for $M_{SB}=100$ GeV. 
We would study these issues in more detail 
in a separate work~\cite{ipgs}. 

\ignore{
\begin{figure}[t]
\begin{center}
\begin{minipage}{0.45\linewidth}
\begin{center}
\epsfig{file=delr21.eps,width=\linewidth}
\end{center}
\end{minipage}
\vfill
\begin{minipage}{0.45\linewidth}
\begin{center}
\epsfig{file=ddll23.eps,width=\linewidth}
\end{center}
\end{minipage}
\begin{minipage}{0.45\linewidth}
\begin{center}
\epsfig{file=ddlr33.eps,width=\linewidth}
\end{center}
\end{minipage}
\end{center}
\caption{The mass insertion parameters 
$(\delta^e_{LR})_{21}$, 
$(\delta^d_{LL})_{23}$ and 
$(\delta^d_{LR})_{33}$ 
as a function of $k \equiv k_1=k_2=k_3$. 
The parameters are chosen as $\phi_X=5$, 
$(\alpha,\beta,\gamma)=(1,0,1)$ and $M_{SB}=100$ GeV.}
\label{deltas}
\end{figure}
}

\section{Summary} \label{summary}
We have studied 4D effective theory of 5D supergravity 
with multi moduli, focusing on 
the contact terms between the hidden and the visible 
hypermultiplets and a resulting flavor structure of the 
soft SUSY breaking terms induced at tree-level. 
The essential difference from the single modulus case 
appears in the K\"{a}hler potential. 

In the single modulus case, the induced soft scalar masses 
by the contact term in the K\"{a}hler potential tend to be 
tachyonic at tree-level. 
This contribution becomes exponentially small when the 
visible sector is geometrically sequestered from 
the SUSY breaking sector. 
In such a case, quantum effects to 
the soft scalar masses become dominant, and may save the 
tachyonic masses at tree-level. 
However, the hierarchical fermion masses by the localized 
wave functions along the extra dimension is incompatible 
with such a sequestering structure~\cite{single_modulus}. 

In the multi moduli case, on the other hand, due to the 
exchange of $Z_2$-odd vector multiplets there is an additional 
contribution to the soft scalar masses that does not suppressed 
even when the quark, lepton multiplets and the SUSY breaking 
multiplet are localized around the opposite boundaries. 
This additional contribution can save the tachyonic scalar 
mass problem in the single modulus case, and can be even 
flavor universal. 
The tree-level contribution to the soft scalar masses always 
dominate over the quantum effects in the multi moduli case. 

Based on these generic features, we constructed a concrete 
model that stabilizes the moduli at a SUSY breaking Minkowski 
minimum where hierarchical Yukawa couplings are generated. 
We have shown the low energy sparticle spectrum of this model 
and analyzed the mass insertion parameters which are relevant 
to some observables of flavor changing neutral currents. 
We should stress that, in our model, all the nontrivial 
structures at low energies are generated dynamically from 
the parameters of ${\cal O}(1)$ (in the Planck unit), 
such as the coefficients of the norm function for the 
vector multiplets, the charges associated with 
the hypermultiplet gauging and the boundary induced 
holomorphic Yukawa couplings. 

There are a lot of directions to proceed based on our work. 
It would be interesting to study models with different 
parameter choices from those we have chosen in this paper~\cite{ipgs}. 
We may also extend the following models to the case with multi moduli,  
e.g., the models with 
two compensator hypermultiplets~\cite{Abe:2007}, 
a twisted $SU(2)_U$ gauge fixing~\cite{Abe:2005wn}, 
an anomalous $U(1)$ symmetry~\cite{Choi:2006bh}, 
moduli mixing nonperturbative effects~\cite{Abe:2005rx} 
and the one in which both the moduli $T^1$ and $T^2$ 
remain dynamical at low energies~\cite{Abe:2005pi}, and so on. 
Another direction is to consider higher dimensional supergravity 
than 5 dimensional, e.g., 
magnetised extra dimensions~\cite{Cremades:2004wa}, 
magnetised orbifold models~\cite{Abe:2008fi} 
which can realize certain localized wavefunctions 
for the matter fields with a more fundamental origin 
of the model, \ie, the string theory. 

Although our 5D model has no correspondence with a certain string 
compactification known until now, the study of 
our simple model would be helpful to understand some basic 
nature of all the models where the physics beyond (M)SSM 
is governed by the dynamics in extra dimensions. 
Of course, our model itself provides a concrete and 
dynamical example of the physics beyond the standard model.

\subsection*{Acknowledgements}
The authors would like to thank Tatsuo~Kobayashi for useful comments. 
This work was supported in part by the Japan Society 
for the Promotion of Science for Young Scientists No.182496 (H.~A.), 
and by the Special Postdoctoral Researchers Program at RIKEN (Y.~S.).

\appendix

\section{$\bdm{Z_2}$-odd part of 5D Weyl multiplet} \label{Z2odd_mul}
In supergravity, coordinate derivatives are covariantized 
for the local SUSY transformation by the gravitino. 
The 4D derivatives~$\der_\mu$ appearing in (\ref{eq:lbulk}) 
are covariantized by the $Z_2$-even gravitino~$\psi_{\mu{\rm R}}^{i=1}$ 
in the $N=1$ Weyl multiplet~$E_W$ 
when we promote the $d^4\tht$ and $d^2\tht$ integrals to 
the $D$- and $F$-term action formulae of the $N=1$ conformal 
SUGRA formulation~\cite{4Doffshell}. 
Here the index~$i$ denotes the doublet index for $\SUu$. 
On the other hand, the derivative~$\der_y$ explicitly appearing in 
the superspace action should be covariantized 
by an $N=1$ multiplet which contains $\psi_{y{\rm R}}^{i=1}$. 

Let us first consider $\der_y$ appearing in the third line 
in (\ref{eq:lbulk}). 
As mentioned below Eq.(52) in Ref.~\cite{Paccetti Correia:2004ri}, 
it should be promoted to 
the ``covariant derivative''~$\hat{\der}_y$ written as 
\be
 \hat{\der}_y \equiv \der_y+\Psi^\alp D_\alp+\Xi^\mu\der_\mu, 
 \label{def_der_y1}
\ee
where $\Psi^\alp$ and $\Xi^\mu$ are $N=1$ superfields with 
4D spinor and vector indices. 
In order for $\hat{\der}_y\Phi^{\hat{a}}$ to be a chiral superfield, 
$\Psi^\alp$ and $\Xi^\mu$ should satisfy the conditions: 
\be
 \bar{D}_{\dot{\alp}}\Psi^\alp=0, \;\;\;
 \Psi^\alp = \frac{i}{8}\bar{D}_{\dot{\alp}}
 \Xi^\mu\bar{\sgm}_\mu^{\dot{\alp}\alp}. 
\ee
The solution to these conditions is 
\be
 \Psi^\alp = \bar{D}^2L^\alp, \;\;\;
 \Xi^\mu = -4i\sgm_{\alp\dot{\alp}}^\mu\bar{D}^{\dot{\alp}}L^\alp, 
 \label{expr_by_L}
\ee
where $L^\alp$ is a complex general multiplet with a spinor index, 
which contains the $Z_2$-odd components of the 5D Weyl multiplet. 
Since the above solution has an ambiguity of adding 
a chiral superfield~$\Omg^\mu$ to $\Xi^\mu$, we can devide $\Xi^\mu$ 
into two part as 
\be
 \Xi^\mu = \Psi^\mu+\Omg^\mu, 
\ee
so that $\Psi^\mu$ contains the same component fields as $\Psi^\alp$. 
The lowest component of $\Psi^\alp$ is identified with 
$\psi_{y{\rm R}}^{i=1\,\alp}$. 

In order for the Lagrangian~(\ref{eq:lbulk}) to be invariant 
under the gauge transformation:~$\Phi^{\hat{a}}\to 
(e^{2ig\Lmd^It_I})^{\hat{a}}_{\;\;\hat{b}}\Phi^{\hat{b}}$, 
we have to modify the gauge transformation of $\Sgm^I$ as 
\be
 \Sgm^I \to \Sgm^I+\hat{\der}_y\Lmd^I, 
\ee
while that of $V^I$ remains unchanged as $V^I\to V^I+\Lmd^I+\bar{\Lmd}^I$. 
Thus the gauge invariant quantity~$\cV^I$ defined below (\ref{def_cN}) 
should also be modified. 
The naive modification is 
\be
 \cV^I \equiv -\hat{\der}_y V^I+\Sgm^I+\bar{\Sgm}^I. 
 \label{def_cV}
\ee
However this is not real nor gauge invariant. 
So we further modify the definition of $\hat{\der}_y$ as 
\be
 \hat{\der}_y \equiv \der_y+\Psi^\alp D_\alp
 +\bar{\Psi}_{\dot{\alp}}\bar{D}^{\dot{\alp}}
 +\frac{i}{4}\Xi^\mu\bar{\sgm}_\mu^{\dot{\alp}\alp}
 \bar{D}_{\dot{\alp}}D_\alp
 +\frac{i}{4}\bar{\Xi}^\mu\sgm_{\mu\alp\dot{\alp}}D^\alp\bar{D}^{\dot{\alp}}, 
 \label{def_der_y2}
\ee
which reduces to the previous definition~(\ref{def_der_y1}) 
when it operates on a chiral superfield. 
With this definition of $\hat{\der}_y$, 
the quantity~$\cV^I$ in (\ref{def_cV}) is now real and gauge invariant. 

Now we obtain the couplings of the $Z_2$-odd part of the 5D Weyl 
multiplet to the matter fields by replacing $\der_y$ explicitly appearing 
in (\ref{eq:lbulk}) with $\hat{\der}_y$ defined in (\ref{def_der_y2}). 
In fact, the terms involving $L^\alp$ are necessary 
for reproducing the correct coefficient functions of 
the kinetic terms for the gauge fields, $-(1/2)(\cN_{IJ}-\cN_I\cN_J/\cN)$, 
where the arguments of $\cN$'s are 
the real scalar components of the 5D vector multiplets. 
If $\der_y$ is not promoted to $\hat{\der}_y$ in the 5D action, 
the reproduced coefficient functions become incorrect ones, $-(1/2)\cN_{IJ}$, 
as mentioned in Appendix~B of Ref.~\cite{Abe:2004}.\footnote{
This discrepancy does not cause a problem in the derivation 
of the 4D effective action because $\cN_I$ are $Z_2$-odd 
for the $Z_2$-even gauge fields and are dropped 
in the procedure of the off-shell dimensional reduction~\cite{Abe:2006eg}. 
} 
As a further nontrivial cross-check, 
we can also see that the couplings of $\Omg_\mu$ to the matter multiplets 
reproduce the correct matter couplings of $V_\mu^{(1)}+iV_\mu^{(2)}$, 
which are the $Z_2$-odd components of the $\SUu$ (auxiliary) gauge field, 
if the $F$-term of $\Omg_\mu$ is identified 
as $2i(V_\mu^{(1)}+iV_\mu^{(2)})$. 

As we can see from (\ref{def_der_y2}) with (\ref{expr_by_L}), 
the $Z_2$-odd part of the 5D Weyl multiplet couples to the matter 
multiplets only in the derivative forms. 
Thus we can neglect their effects on the low-energy effective theory.

\section{Derivation of effective K\"{a}hler potential} \label{eff_Kahler}
Here we explain the derivation of 
the K\"{a}hler potential in the 4D effective theory 
shown in (\ref{eff_def_fcn}) with (\ref{cf_Omg}). 
From (\ref{L_4D:1}) with (\ref{brkt:QS}), 
the effective K\"{a}hler potential is written as 
\be
 \Omg \equiv -3e^{-K/3} = -3\int_0^{\pi R-\ep}\dr y\;\hat{\cN}^{1/3}
 \brkt{1-\sum_ie^{2c_i\tl{V}^1}\abs{Q_i}^2
 -\sum_\alp e^{2c_\alp\tl{V}^2}\abs{S_\alp}^2}^{2/3},  
 \label{expr_Omg}
\ee
where the arguments of $\hat{\cN}$ is  
$(-\der_y\tl{V}^1,-\der_y\tl{V}^2)$. 
The equation of motion for $\tl{V}^1$ are read off from (\ref{L_4D:1}) as 
\bea
 &&\hat{\cN}^{1/3}\frac{\sum_i 4c_ie^{2c_i\tl{V}^1}\abs{Q_i}^2}
 {\brkt{1-\sum_ie^{2c_i\tl{V}^1}\abs{Q_i}^2-\sum_\alp e^{2c_\alp\tl{V}^2}
 \abs{S_\alp}^2}^{1/3}} \nonumber\\
 &&-\der_y\brc{\frac{\hat{\cN}_1}{\hat{\cN}^{2/3}}
 \brkt{1-\sum_ie^{2c_i\tl{V}^1}\abs{Q_i}^2-\sum_\alp e^{2c_\alp\tl{V}^2}
 \abs{S_\alp}^2}^{2/3}} = 0. 
 \label{EOM_tlV}
\eea
In the absence of the matter multiplets~$Q_i$ and $S_\alp$, 
the above equation is reduced to 
\be
 \der_y\brkt{\frac{\hat{\cN}_1}{\hat{\cN}^{2/3}}} = 0. 
 \label{simple_EOM}
\ee
Note that $\hat{\cN}_{I'}/\hat{\cN}^{2/3}$ ($I'=1,2$) depends 
only on the ratio~$v\equiv\der_y\tl{V}^2/\der_y\tl{V}^1$, \ie, 
\be
 \cF_{I'}(v) \equiv \frac{\hat{\cN}_{I'}}{\hat{\cN}^{2/3}}. 
\ee
Thus (\ref{simple_EOM}) means that $v\equiv\bar{v}$, 
where $\bar{v}$ is independent of $y$. 
Then, from the definition of $v$, we obtain a relation:
\be
 \der_y\tl{V}^2 = \bar{v}\der_y\tl{V}^1. 
\ee
By integrating this for $y$ over $[0,\pi R)$, the quantity~$\bar{v}$ 
is determined as  
\be
 \bar{v} = \frac{\Re T^2}{\Re T^1}. 
\ee
We have used (\ref{BC_V:pi}) and (\ref{BC_V:0}). 
In the presence of $Q_i$ and $S_\alp$, the ratio~$v$ is 
expanded in terms of them as 
\bea
 v \eql \bar{v}+\sum_iA_i\abs{Q_i}^2+\sum_\alp B_\alp\abs{S_\alp}^2
 +\sum_{i,j}C_{ij}\abs{Q_i}^2\abs{Q_j}^2 \nonumber\\
 &&+\sum_{i,\alp}D_{i\alp}\abs{Q_i}^2\abs{S_\alp}^2
 +\sum_{\alp,\bt}E_{\alp\bt}\abs{S_\alp}^2\abs{S_\bt}^2
 +\cO(\hat{\Phi}^6), 
 \label{expand_v}
\eea
where $\hat{\Phi}=Q_i,S_\alp$.  
The coefficients~$A_i$, $B_\alp$, $C_{ij}$, $D_{i\alp}$ and $E_{\alp\bt}$ 
are independent of $Q_i$ and $S_\alp$, which satisfy 
\bea
 &&\int_0^{\pi R}\dr y\;\left(\sum_iA_i\abs{Q_i}^2+\sum_\alp B_\alp
 \abs{S_\alp}^2+\sum_{i,j}C_{ij}\abs{Q_i}^2\abs{Q_j}^2 \right. \nonumber\\
 &&\hspace{20mm}\left.
 +\sum_{i,\alp}D_{i\alp}\abs{Q_i}^2\abs{S_\alp}^2
 +\sum_{\alp,\bt}E_{\alp\bt}\abs{S_\alp}^2\abs{S_\bt}^2
 \right)\der_y\tl{V}^1 = \cO(\hat{\Phi}^6). 
 \label{cstrt}
\eea

From (\ref{expr_Omg}) and (\ref{expand_v}), we obtain 
\bea
 \Omg \eql 3\int_0^{\pi R-\ep}\dr y\;\hat{\cN}^{1/3}(1,v)\der_y\tl{V}^1
 \brkt{1-\sum_i e^{2c_i\tl{V}^1}\abs{Q_i}^2
 -\sum_\alp e^{2c_\alp\tl{V}^2}\abs{S_\alp}^2}^{2/3}
 \nonumber\\
 \eql -3\hat{\cN}^{1/3}(1,\bar{v})\left\{
 \Re T^1-\sum_i\frac{1-e^{-2c_i\Re T^1}}{3c_i}\abs{Q_i}^2
 -\sum_\alp\frac{1-e^{-2c_\alp\Re T^2}}{3c_\alp\bar{v}}\abs{S_\alp}^2 \right.
 \nonumber\\
 &&\hspace{25mm}
 -\sum_{i,j}\frac{1-e^{-2(c_i+c_j)\Re T^1}}{18(c_i+c_j)}\abs{Q_i}^2\abs{Q_j}^2
 -\sum_{i,\alp}\frac{1-e^{-2(c_i+c_\alp\bar{v})\Re T^1}}{9(c_i+c_\alp\bar{v})}
 \abs{Q_i}^2\abs{S_\alp}^2 \nonumber\\
 &&\hspace{25mm}\left. 
 -\sum_{\alp,\bt}\frac{1-e^{-2(c_\alp+c_\bt)\Re T^2}}{18(c_\alp+c_\bt)\bar{v}}
 \abs{S_\alp}^2\abs{S_\bt}^2 \right\} \nonumber\\
 &&+\sum_{i,j}\int_0^{\pi R}\dr y\;\brc{
 -\cF_2(\bar{v})A_i\der_y\brkt{\frac{e^{2c_j\tl{V}^1}}{3c_j}}
 +\frac{\cF'_2(\bar{v})}{2}A_iA_j\der_y\tl{V}^1}\abs{Q_i}^2\abs{Q_j}^2
 \nonumber\\
 &&+\sum_{i,\alp}\int_0^{\pi R}\dr y\;\left\{
 \cF_1(\bar{v})A_i\der_y\brkt{\frac{e^{2c_\alp\tl{V}^2}}{3c_\alp\bar{v}^2}}
 -\cF_2(\bar{v})B_\alp\der_y\brkt{\frac{e^{2c_i\tl{V}^1}}{3c_i}}
 \right. \nonumber\\
 &&\hspace{25mm}\left.
 +\cF'_2(\bar{v})A_iB_\alp\der_y\tl{V}^1\right\}\abs{Q_i}^2\abs{S_\alp}^2 
 \nonumber\\
 &&+\sum_{\alp,\bt}\int_0^{\pi R}\dr y\;\brc{
 \cF_1(\bar{v})B_\alp\der_y\brkt{\frac{e^{2c_\bt\tl{V}^2}}{3c_\bt\bar{v}^2}}
 +\frac{\cF'_2(\bar{v})}{2}
 B_\alp B_\bt\der_y\tl{V}^1}\abs{S_\alp}^2\abs{S_\bt}^2
 +\cO(\hat{\Phi}^6). 
 \label{expr_Omg2}
\eea
Here we have used the identity, 
\be
 \cF_1(v)+v\cF_2(v) = 3\hat{\cN}^{1/3}(1,v). 
 \label{identity1}
\ee

In order to calculate the remaining integrals in (\ref{expr_Omg2}), 
we will express $A_i$ and $B_\alp$ in terms of $\tl{V}^1$ and $\tl{V}^2$. 
First, we divide them as 
\be
 A_i = A_{i0}+\Dlt A_i, \;\;\;
 B_\alp = B_{\alp 0}+\Dlt B_\alp, 
\ee
where $A_{i0}\equiv A_i(y=0)$ and $B_{\alp 0}\equiv B_\alp(y=0)$. 
In the absence of $Q_i$, (\ref{EOM_tlV}) is reduced to 
\be
 \der_y\brc{\cF_1(v)\brkt{1-\sum_\alp 
 e^{2c_\alp\tl{V}^2}\abs{S_\alp}^2}^{2/3}} = 0, 
\ee
which means that 
\be
 \cF_1(v)\brkt{1-\sum_\alp e^{2c_\alp\tl{V}^2}\abs{S_\alp}^2}^{2/3}
 = \cF_1(v_0)\brkt{1-\sum_\alp\abs{S_\alp}^2}^{2/3}, 
\ee
where $v_0\equiv v(y=0)$. 
Comparing the coefficients of $\abs{S_\alp}^2$ in both sides, we obtain 
\be
 e^{2c_\alp\tl{V}^2} = 1+\frac{3\cF'_1(\bar{v})}{2\cF_1(\bar{v})}
 \Dlt B_\alp+\cO(\abs{S_\alp}^2). 
\ee
Similar relations are derived for $\tl{V}^1$ and $\Dlt A_i$. 
In the presence of both $Q_i$ and $S_\alp$, they are modified as 
\bea
 \Dlt A_i \eql \frac{2\cF_2(\bar{v})}{3\cF'_2(\bar{v})}
 \brkt{e^{2c_i\tl{V}^1}-1}+\cO(\abs{Q_i}^2,\abs{S_\alp}^2), \nonumber\\
 \Dlt B_\alp \eql \frac{2\cF_1(\bar{v})}{3\cF'_1(\bar{v})}
 \brkt{e^{2c_\alp\tl{V}^2}-1}+\cO(\abs{Q_i}^2,\abs{S_\alp}^2). 
 \label{DltAB}
\eea
From (\ref{cstrt}) with these relations, we can determine 
$A_{i0}$ and $B_{\alp 0}$ as 
\bea
 A_{i0} \eql \frac{2\cF_2(\bar{v})}{3\cF'_2(\bar{v})}
 \brkt{1-\frac{1-e^{-2c_i\Re T^1}}{2c_i\Re T^1}}
 +\cO(\abs{Q_i}^2,\abs{S_\alp}^2), \nonumber\\
 B_{\alp 0} \eql \frac{2\cF_1(\bar{v})}{3\cF'_1(\bar{v})}
 \brkt{1-\frac{1-e^{-2c_\alp\Re T^2}}{2c_\alp\Re T^2}}
 +\cO(\abs{Q_i}^2,\abs{S_\alp}^2). 
 \label{AB0}
\eea

Now we can calculate the remaining integrals in (\ref{expr_Omg2}) 
at the leading order of the $|\hat{\Phi}|^2$-expansion 
by using (\ref{DltAB}) and (\ref{AB0}). 
Since we can use a relation~$\der_y\tl{V}^2=\bar{v}\der\tl{V}^1$ 
in front of the quartic terms in (\ref{expr_Omg2}), 
the integrands can be rewritten as total derivatives for $y$. 
After somewhat lengthy calculations, we obtain the expression of $\Omg$ 
in (\ref{eff_def_fcn}) with (\ref{cf_Omg}). 
Here we have used the following identities. 
\be
 \cF'_1(v)+v\cF'_2(v) = 0, 
\ee
\bea
 \cF'_1(\bar{v}) \eql \Re T^1\der_{\Re T^2}\brkt{
 \frac{\hat{\cN}_1}{\hat{\cN}^{2/3}}}
 = \Re T^1\cdot\frac{3\hat{\cN}\hat{\cN}_{12}-2\hat{\cN}_1\hat{\cN}_2}
 {3\hat{\cN}^{5/3}},  
\eea
\bea
 3\hat{\cN}\hat{\cN}_{12}-2\hat{\cN}_1\hat{\cN}_2 
 \eql -\frac{1}{\bar{v}}\brkt{3\hat{\cN}\hat{\cN}_{11}-2\hat{\cN}_1^2} 
 \nonumber\\
 \eql -\bar{v}\brkt{3\hat{\cN}\hat{\cN}_{22}-2\hat{\cN}_2^2}. 
 \label{identity2}
\eea
The arguments of $\hat{\cN}$, $\hat{\cN_{I'}}$ and $\hat{\cN}_{I'J'}$ 
are $(\Re T^1,\Re T^2)$. 

Finally we comment on a special case in which 
all the gauge couplings~$c_i,c_\alp$ vanish. 
In this case, $\tl{V}^{I'}$ appear in the action~(\ref{L_4D:1}) 
only through $\der_y\tl{V}^{I'}$. 
Then the equation of motion~(\ref{EOM_tlV}) is reduced to (\ref{simple_EOM}), 
which means that $v=\bar{v}$. 
Thus the $y$-integral in (\ref{expr_Omg}) can be easily performed as 
\bea
 \Omg \eql 3\int_0^{\pi R}\dr y\;\der_y\tl{V}^1
 \hat{\cN}^{1/3}(1,\bar{v})
 \brkt{1-\sum_i\abs{Q_i}^2-\sum_\alp\abs{S_\alp}^2}^{2/3} \nonumber\\
 \eql -3\Re T^1\hat{\cN}^{1/3}(1,\bar{v})
 \brkt{1-\sum_i\abs{Q_i}^2-\sum_\alp\abs{S_\alp}^2}^{2/3} \nonumber\\
 \eql -3\hat{\cN}^{1/3}(\Re T)
 \brkt{1-\sum_i\abs{Q_i}^2-\sum_\alp\abs{S_\alp}^2}^{2/3}. 
 \label{simple_Omg}
\eea
Therefore we can obtain the full form of $\Omg$ 
without expanding by $|\hat{\Phi}|^2$ in this case.


\end{document}